\documentclass[%
 reprint,
superscriptaddress,
 amsmath,amssymb,
 aps, article, 
]{revtex4-1}

\usepackage{graphicx}
\usepackage{dcolumn}
\usepackage{bm}
\usepackage{siunitx}
\usepackage{gensymb} 
\usepackage{color} 
\begin{document}

\preprint{}

\title{Critical divergence of the symmetric ($A_{1g}$) nonlinear elastoresistance near the nematic transition in an iron-based superconductor}

 \affiliation{Geballe Laboratory for Advanced Materials and Department of Applied Physics, Stanford University, Stanford, CA 94305, USA.}

 \affiliation{Geballe Laboratory for Advanced Materials and Department of Physics, Stanford University, Stanford, CA 94305, USA.}

\author{J. C. Palmstrom}
\author{A. T. Hristov}

 \affiliation{Geballe Laboratory for Advanced Materials and Department of Applied Physics, Stanford University, Stanford, CA 94305, USA.}

 \affiliation{
 Stanford Institute for Materials and Energy Science, SLAC National Accelerator Laboratory, 2575 Sand Hill Road, Menlo Park, CA  94025, USA.
}

\author{S. A. Kivelson}

 \affiliation{Geballe Laboratory for Advanced Materials and Department of Physics, Stanford University, Stanford, CA 94305, USA.}

 \affiliation{
 Stanford Institute for Materials and Energy Science, SLAC National Accelerator Laboratory, 2575 Sand Hill Road, Menlo Park, CA  94025, USA.
}

\author{J.-H. Chu}
\affiliation{Department of Physics, University of Washington, Seattle WA 98195, USA}
\author{I. R. Fisher}

 \affiliation{Geballe Laboratory for Advanced Materials and Department of Applied Physics, Stanford University, Stanford, CA 94305, USA.}
 \affiliation{
 Stanford Institute for Materials and Energy Science, SLAC National Accelerator Laboratory, 2575 Sand Hill Road, Menlo Park, CA  94025, USA.
}

\date{\today}

\begin{abstract}
We report the observation of a nonlinear elastoresistivity response for the prototypical underdoped iron pnictide Ba(Fe$_{0.975}$Co$_{0.025}$)$_2$As$_2$. Our measurements reveal  a large  quadratic term in the isotropic ($A_{1g}$) electronic response that was produced by a purely shear ($B_{2g}$) strain.  The divergence of this quantity upon cooling towards the structural phase transition reflects the temperature dependence of the nematic susceptibility. This observation shows that nematic fluctuations play a significant role in determining even the isotropic properties of this family of compounds. 

\end{abstract}

\pacs{Valid PACS appear here}
\maketitle

 Nonlinear responses of crystalline materials are described by high rank tensors and can therefore provide valuable information concerning subtle phase transitions and broken symmetries. For example, previous nonlinear measurements of tensor properties have revealed interesting transitions in several strongly correlated materials \cite{hsieh2017, hsieh_cuprates, SH_Rev}. Here we demonstrate a new type of nonlinear transport response, associated with changes in the conductivity of a material in response to strain: nonlinear elastoresistivity. This technique allows us to not only look at broken symmetries across a phase transition, but to characterize properties of the disordered state. We perform these measurements for a representative underdoped Fe-based superconductor, Ba(Fe$_{0.975}$Co$_{0.025}$)$_2$As$_2$, which has previously been shown to exhibit a large nematic susceptibility for temperatures above a tetragonal-to-orthorhombic structural phase transition \cite{Kuo2016, Meingast2016, Uchida2012, Bohmer2014, Forget2013, Hackl2016, chu_2012, kuo2014, kuo2013}. The most remarkable aspect of the current data is that they reveal a diverging nonlinear response in the \emph{isotropic} elastoresistivity in response to a perfectly \emph{antisymmetric} (shear) strain. This observation, which is intimately tied to the large nematic susceptibility of the material studied, serves to underscore the role played by nematic fluctuations in determining even the isotropic properties of the Fe-based superconductors.

Elastoresistivity relates changes in the resistivity ($\Delta\rho=\rho(\epsilon)-\rho(\epsilon=0)$) \footnote{For data shown in this paper, the expansion is done with respect to zero anisotropic strain, $\Delta\rho=\rho(\epsilon)-\rho(\epsilon_{B_{1g/2g}}=0)$. For more details see \cite{EPAPS, Shapiro2015}} to strains ($\epsilon$) experienced by a material; 

\label{eq:der}
\begin{equation}
{
(\frac{\Delta\rho}{\rho_0})_{\alpha}=\sum_{\bar{\alpha}, \bar{\alpha}', ...} (m^{\bar{\alpha}}_{\alpha}\ \epsilon_{\bar{\alpha}}+m^{\bar{\alpha}\bar{\alpha}'}_{\alpha}\ \epsilon_{\bar{\alpha}}\ \epsilon_{\bar{\alpha}'}+...)}
\end{equation}
where the $\alpha$'s  represent a complete, orthogonal basis set for the system, $\epsilon_{\alpha}$ is the component of the overall strain along a given basis vector, and $\rho_0$ is an appropriate normalization factor \cite{Shapiro2015}; here, the in-plane resistivity of the tetragonal phase. A natural basis to work in is the irreducible representations of the crystallographic point group. In the absence of a magnetic field and in the $D_{4h}$ point group (appropriate for the material studied here), both strain and $\Delta\rho/\rho_0$ 
have six independent components. Of these, four unique combinations
correspond to distinct representations: $(\Delta\rho/\rho_0)_{B_{1g}}=\frac{1}{2}[(\Delta\rho/\rho_0)_{xx}-(\Delta\rho/\rho_0)_{yy}]$, $(\Delta\rho/\rho_0)_{B_{2g}} = (\Delta\rho/\rho_0)_{xy}$, and $(\Delta\rho/\rho_0)_{E_{g}} = ((\Delta\rho/\rho_0)_{xz}, (\Delta\rho/\rho_0)_{yz})$. Objects with $B_{1g}$ and $B_{2g}$ symmetry are antisymmetric (odd) with respect to a 90$^o$ rotation about the $z$-axis. There is also a two-dimensional space of components belonging to the $A_{1g}$ representation, the basis of which is not uniquely defined by symmetry alone \footnote{A standard delineation for the two $A_{1g}$ symmetry terms in $\epsilon$ and $\Delta\rho/\rho_0$, which we also adopt here, is to separate in-plane ($\alpha = A_{1g,1}$ (i.e. $x^2 + y^2$)) and out-of-plane responses ($\alpha = A_{1g,2}$ (i.e. $z^2$))}. Objects with $A_{1g}$ symmetry are symmetric (even) with respect to a 90$^o$ rotation around the $z$-axis. In this paper we focus on one (of the two) components with $A_{1g}$ symmetry reflecting the in-plane changes in resistivity i.e. $(\Delta\rho/\rho_0)_{A_{1g}}=\frac{1}{2}[(\Delta\rho/\rho_0)_{xx}+(\Delta\rho/\rho_0)_{yy}]$.

The linear elastoresistivity response is described by a fourth rank tensor, which in the present basis corresponds to $m^{\bar\alpha}_{\alpha}$. As shown previously, $m^{B_{1g}}_{B_{1g}}$ and $m^{B_{2g}}_{B_{2g}}$ \footnote{ The irreducible representation notation can be related to the Cartesian and Voigt notations that have been used previously: $m^{B_{1g}}_{B_{1g}} = m_{xx,xx}-m_{xx,yy} = m_{11}-m_{12}$ and $m^{B_{2g}}_{B_{2g}} = 2m_{xy,xy} = 2m_{66}$} are proportional to the nematic susceptibility in the corresponding symmetry channels, $\chi_{B_{1g}}$ and $\chi_{B_{2g}}$ \cite{Shapiro2015, kuo2014, Kuo2016, chu_2012, kuo2013}. To linear order, correctly decomposed symmetry channels cannot mix. For example, for a tetragonal material, antisymmetric strain ($\epsilon_{B_{1g}}$ and $\epsilon_{B_{2g}}$) cannot cause a symmetric resistivity response, 
i.e. $m^{B_{1g}}_{A_{1g}}=m^{B_{2g}}_{A_{1g}}=0$.
However, this is not true when considering the nonlinear response. In the present work, we demonstrate the presence of a large and strongly temperature dependent nonlinear $A_{1g}$ elastoresistivity in response to antisymmetric $B_{2g}$ strain (i.e. we show that $m_{A_{1g}}^{B_{2g}, B_{2g}}\gg1$). We further show that this behavior reflects the diverging nematic susceptibility of the material. 

Measuring the elastoresistance in the $A_{1g}$ symmetry channel presents several technical challenges.  In order to precisely decompose the elastoresistance response into the isotropic 
 and antisymmetric components, the resistivity in two orthogonal directions must be measured simultaneously \emph{for identical strain conditions}; otherwise, the $B_{2g}$ elastoresistance (which for these materials is much larger than the $A_{1g}$ elastoresistivity response) gets admixed. A second important consideration is that to confidently extract the linear and quadratic $A_{1g}$ elastoresistance coefficients, the sample must be close to conditions of neutral anisotropic strain ($\epsilon_{x'x'} - \epsilon_{y'y'} \approx 0$; here the primed coordinate frame refers to the normal strain frame \cite{Shapiro2015}). As we demonstrate, a modified Montgomery technique is especially suitable for both purposes \cite{Kuo2016}. The crystals are cut into thin square plates with the electrical contacts made at the four corners, enabling measurement of $\rho_{x'x'}$ and $\rho_{y'y'}$ simultaneously while the crystal is held under a measured set of strain conditions. The $B_{2g}$ neutral strain point is determined by the condition of $\rho_{x'x'} = \rho_{y'y'}$, since for a crystal with tetragonal symmetry the in-plane resistivity is isotropic if there is zero anisotropic strain. Results for an alternative experimental protocol based on a transverse resistance measurement \cite{Shapiro2016a} are in broad agreement and are presented in the supplemental material. 

\begin{figure}[!h]  
\begin{center}  
\includegraphics[width=3.3in]{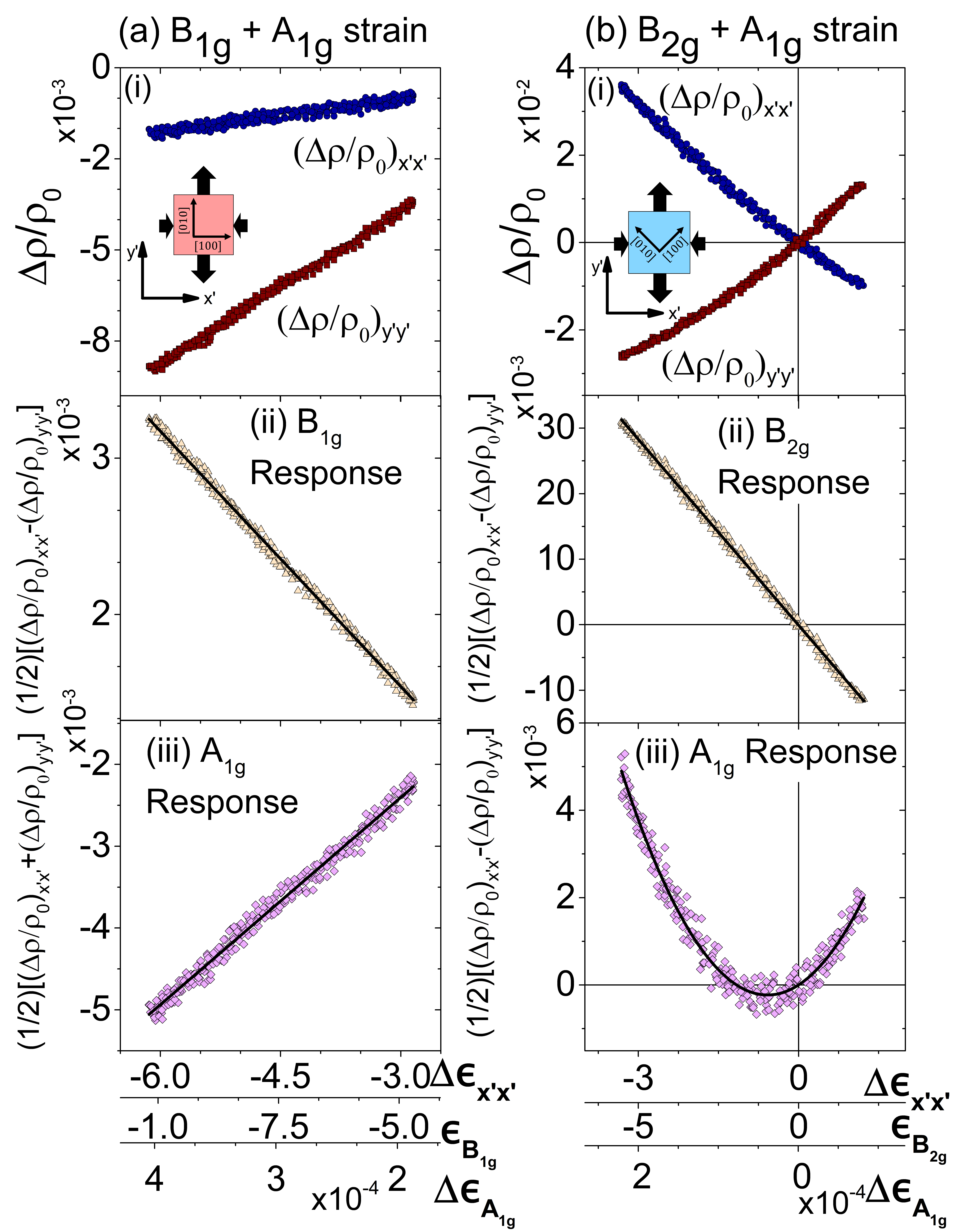}
\caption{\small \sl Representative data showing the resistivity response to strain of Ba(Fe$_{0.975}$Co$_{0.025}$)$_{2}$As$_2$ at 116 K. The left-hand column (a) shows data for a crystal oriented with the crystal axes parallel to the normal strain frame (represented by the schematic pink-colored crystal in the inset to panel (a)(i)), such that the crystal experiences an admixture of $A_{1g}$ and $B_{1g}$ symmetry strain. The right-hand column (b) shows data for a crystal with the axes oriented at 45 degrees to the normal strain frame (shown schematically by the blue crystal in the inset to panel (b)(i)), such that the crystal experiences an admixture of $A_{1g}$ and $B_{2g}$ symmetry strain. The top graph (i) in each column shows the resistive response of the sample along the $x'$ and $y'$ axes due to the strain, where the $x'$ and $y'$ axes are defined by the normal strain frame (inset). The zero antisymmetric strain condition is marked by a vertical line in panel (b). The middle graph (ii) shows the antisymmetric response, given by the difference $\frac{1}{2}[(\Delta\rho/\rho_0)_{x'x'} - (\Delta\rho/\rho_0)_{y'y'}]=(\frac{\Delta\rho}{\rho_0})_{B_{1g}/B_{2g}}$. For both crystal orientations, the antisymmetric response is linear (black lines show linear fits). The bottom graph (iii) shows the symmetric ($A_{1g}$) response, given by the sum $\frac{1}{2}[(\Delta\rho/\rho_0)_{x'x'}$ + $(\Delta\rho/\rho_0)_{y'y'}]=(\frac{\Delta\rho}{\rho_0})_{A_{1g}}$. This response is found to be always linear for samples that experience $A_{1g} + B_{1g}$ symmetry strain (black line shows linear fit), while that of the samples that experience $A_{1g} + B_{2g}$ symmetry strain is clearly nonlinear and is fit by a second order polynomial (black line).}

\label{fig:Sweep}  
\end{center}  
\end{figure}

  In our experimental setup, we apply biaxial stress to the samples by affixing them to a lead-zirconate-titanate (PZT) stack (Part No.: PSt150/5x5/7 cryo 1, from Piezomechanik GmbH). When positive voltage is applied to the PZT stack, it expands along its poling axis (the $y'$ axis) and contracts along the perpendicular axis (the $x'$ axis). For thin samples, the crystal deforms with the PZT stack. The ratio of the strain experienced by the sample along the $y'$ and $x'$ axes is dictated by the in-plane Poisson ratio, $\nu_P$, of the PZT stack ($\epsilon_{y'y'}=-\nu_P\epsilon_{x'x'}$). This is a weakly temperature dependent quantity, with an average value for our PZT stacks of $\sim 2.3$. Since the magnitude of strains along the $x'$ and $y'$ directions are not equal, the strain can be decomposed into two parts: a part that is even with respect to rotation by 90$^o$ about the $z$-axis (in-plane $A_{1g}$ symmetry; $\epsilon_{A_{1g}} = \frac{1}{2}(\epsilon_{x'x'} + \epsilon_{y'y'}))$, and an odd part ($B_{1g/2g}$ symmetry; $\epsilon_{B_{1g/2g}} = \frac{1}{2}(\epsilon_{x'x'} - \epsilon_{y'y'})$). As shown in the inset of Fig. \ref{fig:Sweep}, by aligning the sample’s square edges along either the tetragonal [100] or tetragonal [110] direction, we selectively cause the material to experience $A_{1g} + B_{1g}$ symmetry strain (pink) or $A_{1g} + B_{2g}$ symmetry strain (blue). More experimental details can be found in the supplemental material \cite{EPAPS}.

There is a qualitative difference in the strain-dependence of the elastoresistivity between samples that experience $B_{1g}$ and $B_{2g}$ symmetry strain. Fig. \ref{fig:Sweep} shows representative data for Ba(Fe$_{0.975}$Co$_{0.025}$)$_{2}$As$_2$ above the structural phase transition.
 Multiple samples of both orientations have been measured \cite{EPAPS}. The sample that experiences $B_{1g}$ strain exhibits a linear change in $\rho_{x'x'}$ and $\rho_{y'y'}$ under strain. Consequently, both the antisymmetric response ($(\Delta\rho/\rho_0)_{B_{1g}}$) and the symmetric response ($(\Delta\rho/\rho_0)_{A_{1g}}$) are also linear in strain.  In contrast, the sample that experiences $B_{2g}$ strain exhibits a clear nonlinearity in both  $\rho_{x'x'}$ and $\rho_{y'y'}$ as the strain is varied. The antisymmetric ($B_{2g}$) response  is perfectly linear (black line in Fig. \ref{fig:Sweep}(b)(ii)) and comparatively large, whereas the symmetric ($A_{1g}$) response exhibits a striking nonlinearity and is fit by a quadratic function (black line in Fig. \ref{fig:Sweep}(b)(iii)). The minimum of the quadratic function does not occur at the same strain as the neutral $B_{2g}$ strain point (vertical line in Fig. \ref{fig:Sweep}(b)), indicating the presence of a linear term in addition to the quadratic coefficient.

\begin{figure}[!htbp] 
\begin{center}  
\includegraphics[width=3.4in]{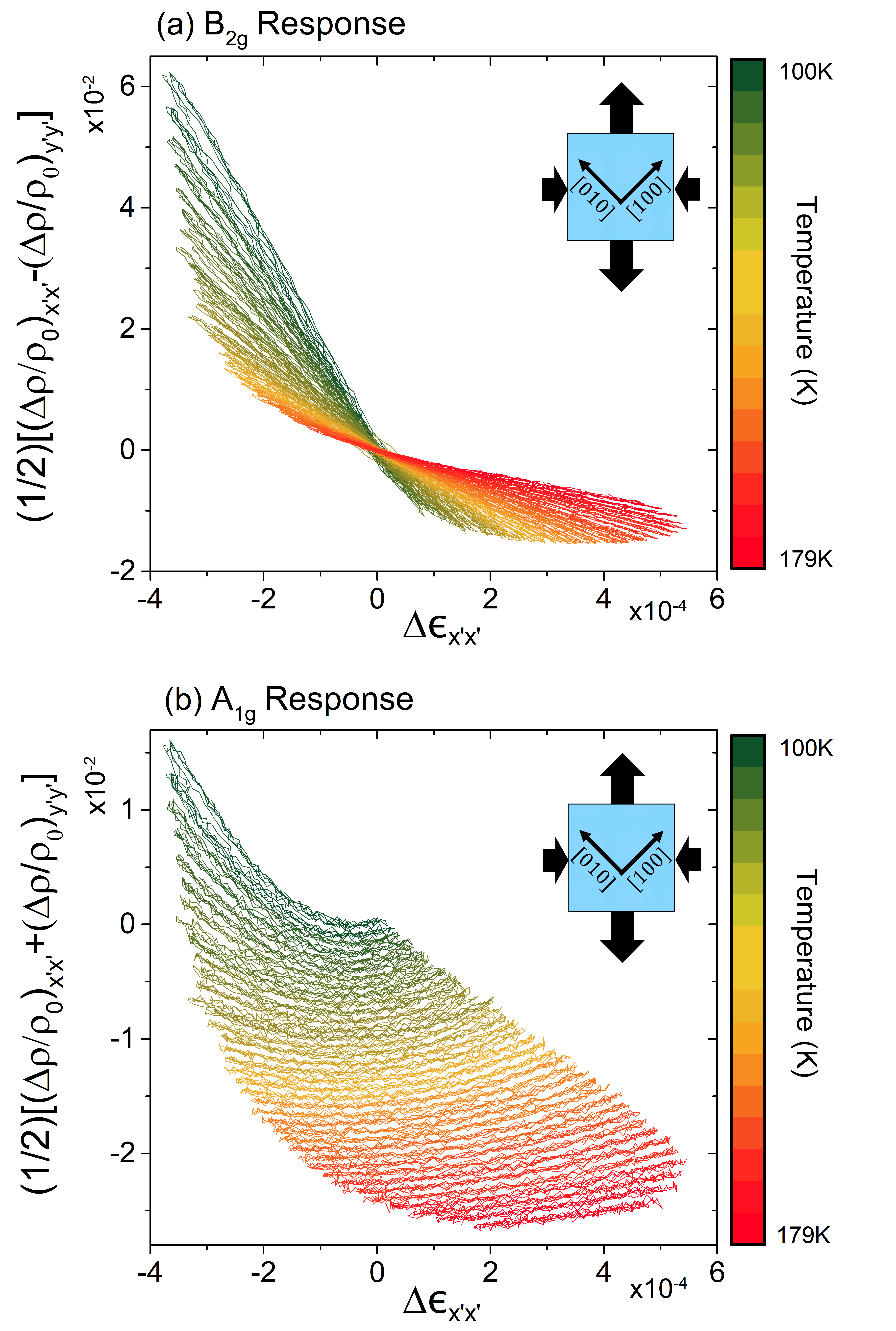}
\caption{{ \small \sl Temperature dependence of (a) the antisymmetric ($B_{2g}$) elastoresistivity response, and (b) the isotropic ($A_{1g}$) elastoresistivity response, of a single crystal of Ba(Fe$_{0.975}$Co$_{0.025})_{2}$As$_2$ oriented with the crystal axes at 45 degrees to the normal strain frame (blue schematic insets). The anisotropic response is always linear, whereas the isotropic response shows a large quadratic component with a minimum close to the $B_{2g}$ neutral strain point. Both responses exhibit a strong temperature dependence. Note that the accessible strain range shifts with temperature, due in part to differences in the thermal expansion of the PZT and sample, and in part to the temperature dependence of the dynamic range of the PZT stack. For clarity, each fixed temperature strain sweep for the $A_{1g}$ response are offset by $-7.5\times10^{-4}$ per trace from the 100K sweep. The data showing the $B_{2g}$ response are not offset.}}
\label{fig:waterfall}  
\end{center}  
\end{figure}

The qualitative behavior shown in Fig. \ref{fig:Sweep} is characteristic of both crystal orientations for the range of measured temperatures. Data of the elastoresistance response at different temperatures are shown in Fig. \ref{fig:waterfall} for the sample that was oriented to experience $B_{2g}$ symmetry strain; similar data for $B_{1g}$ symmetry strain are shown in the supplemental material \cite{EPAPS}. For $B_{2g}$ symmetry strains, the antisymmetric response is linear for all temperatures measured, with a slope that grows larger as temperature decreases. Similarly, the symmetric ($A_{1g}$) response exhibits a strong temperature dependence, with a clear increase in the coefficient of the quadratic term as temperature is reduced towards the structural transition. In contrast, the sample that experiences $B_{1g}$ symmetry strain exhibits only a weak temperature dependence in the linear response for both symmetry channels, as shown in Fig \ref{fig:Sweep}(a), and never exhibits any measurable nonlinearity. 

\begin{figure}[!ht]  
\begin{center}  
\includegraphics[width=2.8in]{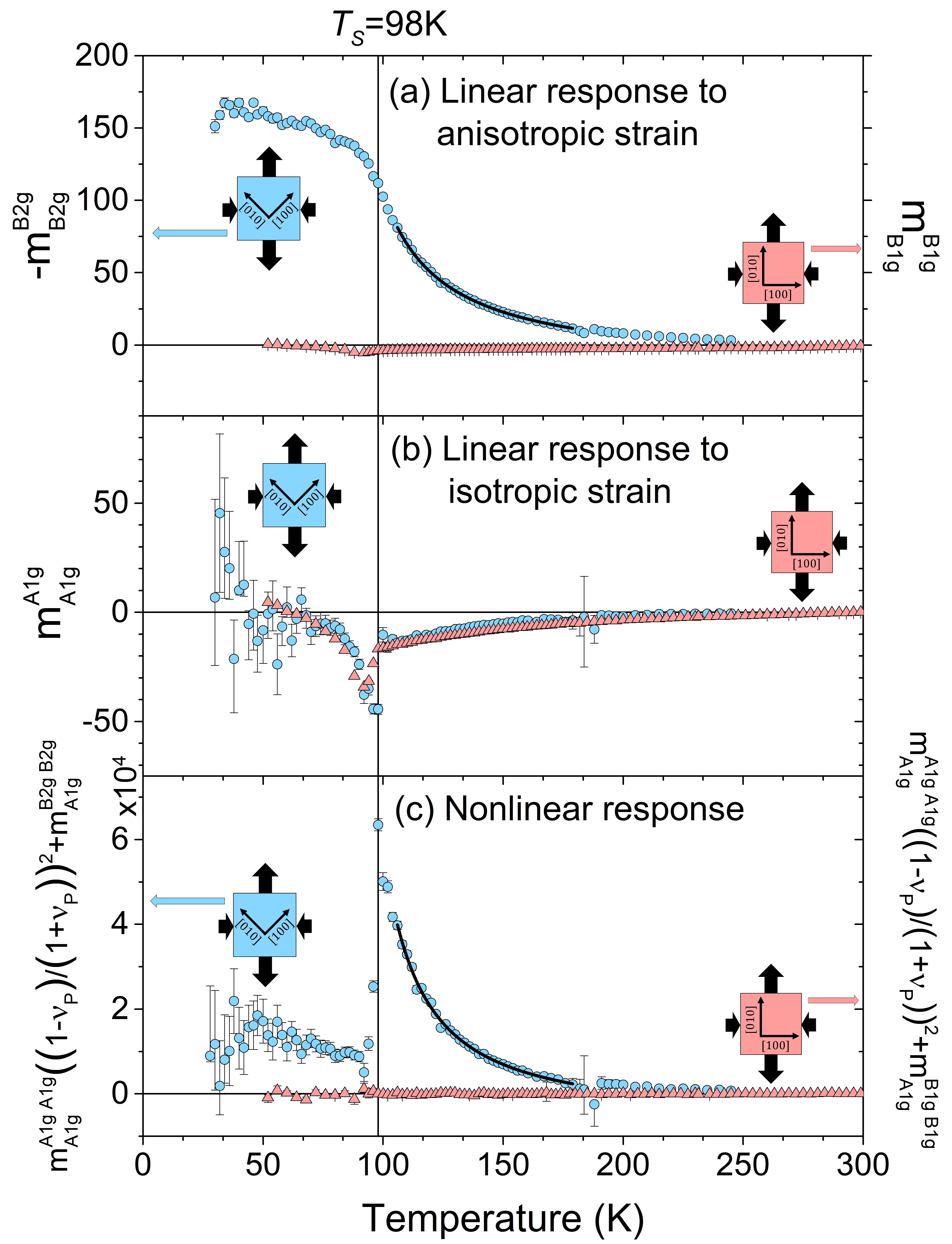}
\caption{\small \sl Temperature dependence of the elastoresistance coefficients of Ba(Fe$_{0.975}$Co$_{0.025}$)$_{2}$As$_2$ for all symmetry channels measured. Blue circles show the response for a sample that experiences an admixture of $A_{1g} + B_{2g}$ symmetry strain (blue schematic insets), while pink triangles show the response for a sample that experiences an admixture of $A_{1g} + B_{1g}$ symmetry strain (pink schematic insets). (a) The linear response to anisotropic strain, $m^{B_{2g}}_{B_{2g}}$ (left axis) and $m^{B_{1g}}_{B_{1g}}$ (right axis). $m^{B_{2g}}_{B_{2g}}$ can be well fit by a Curie-Weiss functional form (black line; see main text). (b) The linear response to isotropic strain, $m^{A_{1g}}_{A_{1g}}$. For crystals that experience $A_{1g} + B_{1g}$ symmetry strain (pink), $m^{A_{1g}}_{A_{1g}}$ is extracted from a linear fit; for crystals that experience $A_{1g} + B_{2g}$ symmetry strain (blue), the data are extracted from the linear term of a second order polynomial fit.  (c) The weighted quadratic coefficients, $((\frac{1-\nu_P}{1+\nu_P})^2m_{A_{1g}}^{A_{1g}, A_{1g}} + m_{A_{1g}}^{B_{2g}, B_{2g}})$ (blue data, left axis) and $((\frac{1-\nu_P}{1+\nu_P})^2m_{A_{1g}}^{A_{1g}, A_{1g}} + m_{A_{1g}}^{B_{1g}, B_{1g}})$ (pink data, right axis) describing the isotropic response to ($A_{1g} + B_{2g}$) and ($A_{1g} + B_{1g}$) symmetry strains, respectively, extracted from the 2nd order polynomial fit of the isotropic response as a function of anisotropic strain. The only measurably non-zero non-linear coefficient is $m_{A_{1g}}^{B_{2g}, B_{2g}}$, the isotropic response to $B_{2g}$ symmetry anisotropic strain. The temperature-dependence of this coefficient can be well fit by $\frac{a}{(T-\Theta)^2} + \frac{b}{T-\Theta} + c$ (black line; see main text), with $\Theta$ taken from the Curie Weiss fit to $m^{B_{2g}}_{B_{2g}}$. Error bars represent 95\% confidence intervals from statistical fits. If an error bar is not shown, the uncertainty of the fit is contained within the size of the data point.}  
\label{fig:summary}  
\end{center}  
\end{figure}

We first consider the linear response to antisymmetric strains, $m^{B_{1g}}_{B_{1g}}$ and $m^{B_{2g}}_{B_{2g}}$, shown in Fig. \ref{fig:summary}(a). As found previously \cite{kuo2013}, $m^{B_{1g}}_{B_{1g}}$ is small and exhibits almost no temperature dependence. In contrast, $m^{B_{2g}}_{B_{2g}}$ is large and can be well fit by a Curie-Weiss temperature dependence with a Weiss temperature $\Theta$ = 75.8 $\pm$ 0.6 K (adjusted R-squared, $R^2_{adj} = 0.9995$), bearing witness to the divergent nematic susceptibility in this material \cite{chu_2012,kuo2013, kuo2014, Kuo2016}. The coupled nematic/structural phase transition occurs at a higher temperature $T_s = 98$ $\pm$ 2 K due to bilinear coupling between the nematic order parameter and lattice strain with the same symmetry \cite{chu_2012}. 

The linear response to $A_{1g}$ strain, $m^{A_{1g}}_{A_{1g}} $, is small and only weakly temperature-dependent (Figure \ref{fig:summary}(b)) \footnote{For the $D_{4h}$ point symmetry, there are four independent combinations of terms in the 4th rank elastoresistivity tensor that can contribute to an $A_{1g}$ symmetry elastoresistance response. Two possible linear combinations of strain that transform like an $A_{1g}$ object are; $\epsilon_{A_{1g,1}} =\frac{1}{2}[\epsilon_{x'x'}+\epsilon_{y'y'}$] and $\epsilon_{A_{1g,2}}=\epsilon_{z'z'}$ and similarly for $\Delta\rho/\rho_0$; $\rho_{A_{1g,1}} = \frac{1}{2}[(\Delta\rho/\rho_0)_{x'x'}$ + $(\Delta\rho/\rho_0)_{y'y'}$] and $\rho_{A_{1g,2}} = (\Delta\rho/\rho_0)_{z'z'}$ \cite{Shapiro2015}. For the measurements described in this paper, the effective out of plane Poisson ratio $\nu_z$ of the sample bonded to the PZT stack determines the strain in the $z$ direction ($\epsilon_{zz} = -\nu_z\epsilon_{xx}$), which in turn affects the degree to which these combinations of coefficients admix in the measured symmetric response. Hence,

\label{eq:sym_irrep}
\begin{equation}
(\frac{\Delta\rho}{\rho_0})_{A_{1g,1}}=m^{A_{1g,1}}_{A_{1g,1}}\epsilon_{A_{1g,1}}+m^{A_{1g,2}}_{A_{1g,1}}\epsilon_{A_{1g,2}}
\end{equation}

Since $\epsilon_{z'z'}=-\nu_z\epsilon_{x'x'}$ and $\epsilon_{y'y'}=-\nu_P\epsilon_{x'x'}$, we obtain

\begin{align}
(\frac{\Delta\rho}{\rho_0})_{A_{1g,1}}&=(m^{A_{1g,1}}_{A_{1g,1}}-\frac{2\nu_z}{1-\nu_P}m^{A_{1g,2}}_{A_{1g,1}})\epsilon_{A_{1g,1}}\\
&=m^{A_{1g}}_{A_{1g}}\epsilon_{A_{1g,1}}
\end{align}

defining the quantity $m^{A_{1g}}_{A_{1g}}$ used in the main text.}. Moreover, values of $m^{A_{1g}}_{A_{1g}}$ determined from both crystal orientations agree (as they must, since by symmetry both $\epsilon_{A_{1g}}$ and $(\Delta\rho/\rho_0)_{A_{1g}}$ are invariant to rotations about the $z$-axis), providing additional confidence that the $B_{2g}$ neutral strain point has been accurately identified.

From a symmetry perspective, non-linear contributions to $(\Delta\rho/\rho_0)_{A_{1g}}$ are possible due to all three strains considered. To quadratic order,

\begin{equation} \label{eq:nonlinear}
\begin{split}
 (\frac {\Delta\rho}{\rho_0})_{A_{1g}} &= m^{A_{1g}}_{A_{1g}}\ \epsilon_{A_{1g}} + m_{A_{1g}}^{A_{1g}, A_{1g}}\ [\epsilon_{A_{1g}}]^2\\ 
&+ m_{A_{1g}}^{B_{1g}, B_{1g}}\ [\epsilon_{B_{1g}}]^2 + m_{A_{1g}}^{B_{2g}, B_{2g}}\ [\epsilon_{B_{2g}}]^2\ 
\end{split}
\end{equation}

Since the symmetric and antisymmetric strains are related via $\nu_P$ (i.e. $\epsilon_{B_{1g/2g}} = \frac{(1+\nu_P)}{(1-\nu_P)}\epsilon_{A_{1g}}$), the quadratic coefficient of $(\Delta\rho/\rho_0)_{A_{1g}}$ as a function of $\epsilon_{B_{1g/2g}}$ is given by the weighted sum of coefficients $m_{A_{1g}}^{B_{1g},B_{1g}/B_{2g}, B_{2g}}+(\frac{1-\nu_P}{1+\nu_P})^2 m_{A_{1g}}^{A_{1g}, A_{1g}}$ for $A_{1g}$ + $B_{1g/2g}$ symmetry strains, respectively. The temperature dependence of these weighted sums, obtained from quadratic fits to the data shown in Fig. \ref{fig:waterfall}(b) with appropriate transformation of the strain axis, are plotted in Fig. \ref{fig:summary}(c). Evidently, $m_{A_{1g}}^{A_{1g},A_{1g}}$ and $m_{A_{1g}}^{B_{1g},B_{1g}}$ (the weighted sum of which is shown by the pink data) are vanishingly small. Hence, the striking non-linear response seen in Fig. \ref{fig:waterfall}(b) derives solely from $m_{A_{1g}}^{B_{2g}, B_{2g}}$; that is, \emph{the non-linear symmetric response derives solely from purely antisymmetric ($B_{2g}$) strain}. 

The Curie-Weiss temperature dependence of $m^{B_{2g}}_{B_{2g}}$ directly attests to the presence of an electronic degree of freedom (the nematic order parameter $\phi_{B_{2g}}$) that is separate from, though bi-linearly coupled to, anisotropic strain $\epsilon_{B_{2g}}$: $\phi_{B_{2g}} = \chi_{B_{2g}}\epsilon_{B_{2g}}\propto m^{B_{2g}}_{B_{2g}}\epsilon_{B_{2g}}$. From the same perspective,  in addition to a bare contribution to  $(\Delta\rho/\rho_0)_{A_{1g}}$ that is directly proportional to $[\epsilon_{B_2g}]^2$, there should be additional induced terms proportional to $\phi_{B_{2g}} \epsilon_{B_{2g}}$ and $[\phi_{B_{2g}}]^2$.  All these terms are allowed by symmetry, and since $\phi_{B_{2g}} = \chi_{B_{2g}}\epsilon_{B_{2g}}$, the latter two contributions should be increasingly strong with decreasing temperature, so that:

\label{eq:nonlinear_coeff}
\begin{equation}
m_{A_{1g}}^{B_{2g}, B_{2g}} 
\approx \frac{a}{(T-\Theta)^2} + \frac{b}{T-\Theta} + c
\end{equation}
where $a$, $b$, and $c$ are coefficients to be determined. The Weiss temperature $\Theta$, which is independently determined from the temperature dependence of $m^{B_{2g}}_{B_{2g}}$, is not a fit parameter. The black line in Fig. \ref{fig:summary}(c) shows the best fit to this functional form, with $\sqrt{a}=4\pm1\times10^3$ K and $b=7\pm1\times10^5$ K; both terms are important and necessary to fully fit the response \cite{EPAPS}. This fit is in excellent agreement with the data ($R^2_{adj} = 0.99655$) and confirms our understanding of the contributing symmetry terms and the underlying physics. The quality of fit also implies that the proportionality constant relating $\chi_{B_{2g}}$ and the elastoresistivity coefficients have negligible temperature dependence over the fit range. 

Finally, we note that $m_{A_{1g}}^{B_{2g},B_{2g}}$ is positive. This implies that the average resistance is expected to be larger in the anisotropic nematic phase than an extrapolation of the in-plane resistivity determined from the isotropic tetragonal state. Since this is a second order effect, we expect the resistivity increase to scale as the square of the on-setting nematic order parameter, i.e. to have a T-linear temperature dependence, for temperatures close to $T_s$. This is consistent with the observation \cite{Chu2009} that the resistivity of twinned Ba(Fe$_{0.975}$Co$_{0.025}$)$_{2}$As$_2$ samples linearly increases upon cooling through the structural transition \footnote{While domain wall scattering can also contribute to the average resistivity of twinned samples in this regime, the linear temperature dependence implies that the dominant effect derives from the intrinsic increase in the average resistivity.}.
   
The most remarkable aspect of this measurement is not that $m_{A_{1g}}^{B_{2g}, B_{2g}}\neq 0$, since this is allowed by symmetry, but how large this quantity is. Indeed, close to the structural transition the nonlinear response of $(\Delta\rho/\rho_0)_{A_{1g}}$ to $\epsilon_{B_{2g}}$ is an order of magnitude larger than the linear response to $\epsilon_{A_{1g}}$ for the range of strain considered here. Furthermore the temperature dependence of this coefficient directly reveals that the effect is driven by the large nematic susceptibility of the material, meaning that even the isotropic properties of the Fe-based superconductors (in this case $(\Delta\rho/\rho_0)_{A_{1g}}$) are strongly affected by the nematic character of the material. These observations demonstrate a new means to witness the divergent nematic susceptibility in these materials based on the measurement of the \emph{isotropic} response to anisotropic strain. They also provide a new point of comparison for microscopic models of the transport properties of Fe-based superconductors.

\setlength{\parskip}{0em}

\section*{ACKNOWLEDGMENTS}

J.C.P. and A.T.H. are supported by a NSF Graduate Research Fellowship (grant DGE-114747). J.C.P. is also supported by a Gabilan Stanford Graduate Fellowship. J.-H. C. acknowledges the support from the State of Washington funded Clean Energy Institute. This work was supported by the Department of Energy, Office of Basic Energy Sciences, under contract no. DE-AC02-76SF00515.

\bibliography{Nonlinear_ER}

\clearpage
\widetext 

\begin{center}
\textbf{\large Supplemental Material: Critical divergence of the symmetric ($A_{1g}$) nonlinear elastoresistance near the nematic transition in an iron-based superconductor}
\end{center}

\makeatletter
\renewcommand{\theequation}{S\arabic{equation}}
\renewcommand{\thetable}{S\arabic{table}}
\renewcommand{\thefigure}{S\arabic{figure}}
\renewcommand{\bibnumfmt}[1]{[S#1]}
\renewcommand{\citenumfont}[1]{S#1}

\tableofcontents

\section{Experimental Methods}\label{sec:method}

\begin{figure}[!h]  
\begin{center}  
\includegraphics[width=4in]{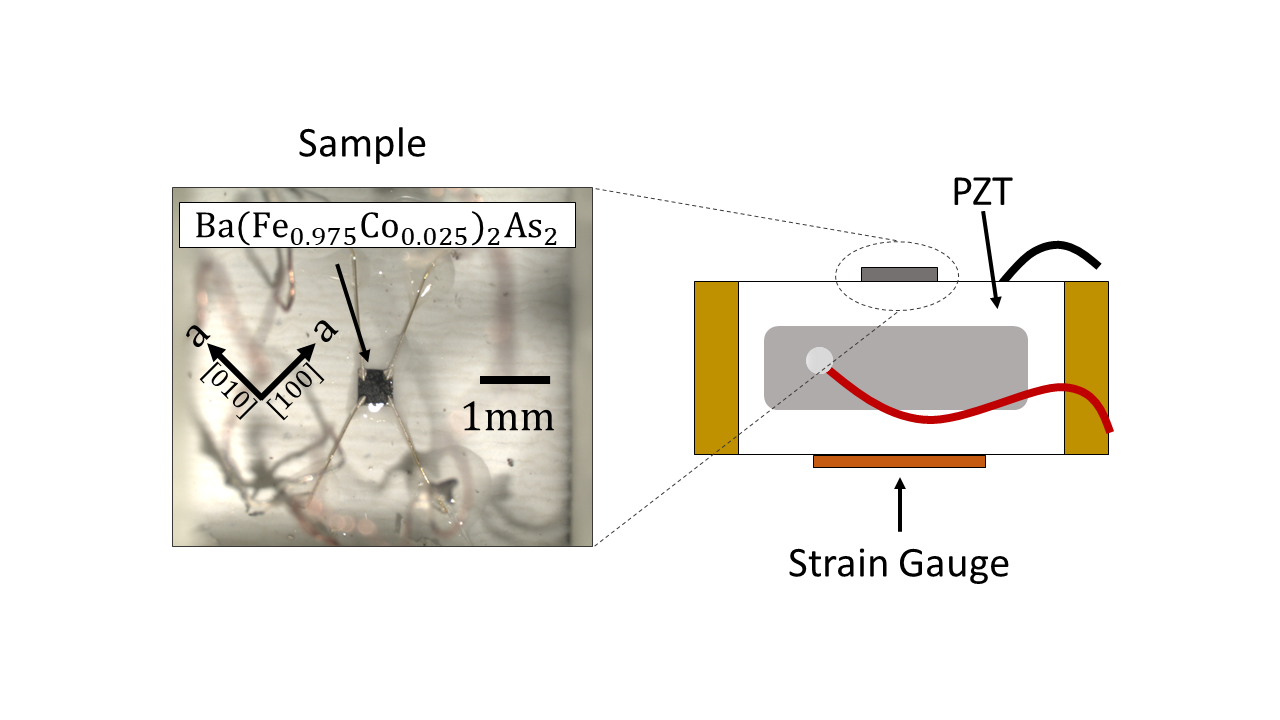}
\caption{\small \sl Left) Photograph of a representative Ba(Fe$_{0.975}$Co$_{0.025}$)$_2$As$_2$ sample ($B_{2g}$ Sample 3) prepared for an elastoresistance measurement using the Modified Montgomery method. The sample is cut into a square shape and glued onto the PZT stack. This sample is aligned with its crystallographic axes rotated 45 degrees with respect to the normal strain frame. Right) Schematic diagram showing the PZT stack prepared for an elastoresistance measurement. The sample is glued to the top face of the PZT.  Strain is measured via a strain gauge glued to the back of the PZT stack.}

\label{fig:method}  
\end{center}  
\end{figure}

The single crystals of Ba(Fe$_{0.975}$Co$_{0.025}$)$_2$As$_2$ were grown using the FeAs self flux technique as described elsewhere \cite{Chu2009}. The crystals were cleaved into thin plates and cut into squares with typical side lengths of 400-750 $\mu$m and thicknesses of 15-30 $\mu$m. These samples were then contacted on the corners of their top surface via sputtered gold pads with gold wires dipped in an air-dry silver epoxy (Dupont 4929N), and glued to the PZT stack with either Devcon 5-minute epoxy or Master Bond EP21TCHT-1. A photograph of a typical sample and a diagram of the PZT setup can be seen in Fig. \ref{fig:method}. Stress was applied to the sample by step wise cycling the voltage from -150 V to 150 V (below 150 K) and -50 V to 150 V (above 150 K) on the PZT stack at a fixed temperature. Three to four voltage sweeps were performed at each temperature, with typical voltage ramp rates between 8-15 V/s. Using the Modified Montgomery Method (MMM) \cite{Kuo2016}, $\rho_{x'x'}$ and $\rho_{y'y'}$ were measured simultaneously at each voltage step. 

The strain is measured by a strain gauge (Part No.: WK-06-062TT-
350 from Micro-Measurements) glued to the back of the PZT stack. Typically only one direction of strain is measured and the orthogonal strain is calculated using the measured Poisson ratio of the PZT stack ($\epsilon_{y'y'}=-\nu_P\epsilon_{x'x'}$) \cite{kuo2013}. For measurements done here we assume perfect strain transmission through the glue and sample. Imperfect strain transmission would scale the resistive response in all symmetry channels, but would neither change our symmetry decomposition nor affect our main conclusions. This is discussed in detail in Sec. \ref{sec:strain}.

\section{Decomposition of the in-plane strain into symmetric and antisymmetric components}

\begin{figure}[!h]  
\begin{center}  
\includegraphics[width=4in]{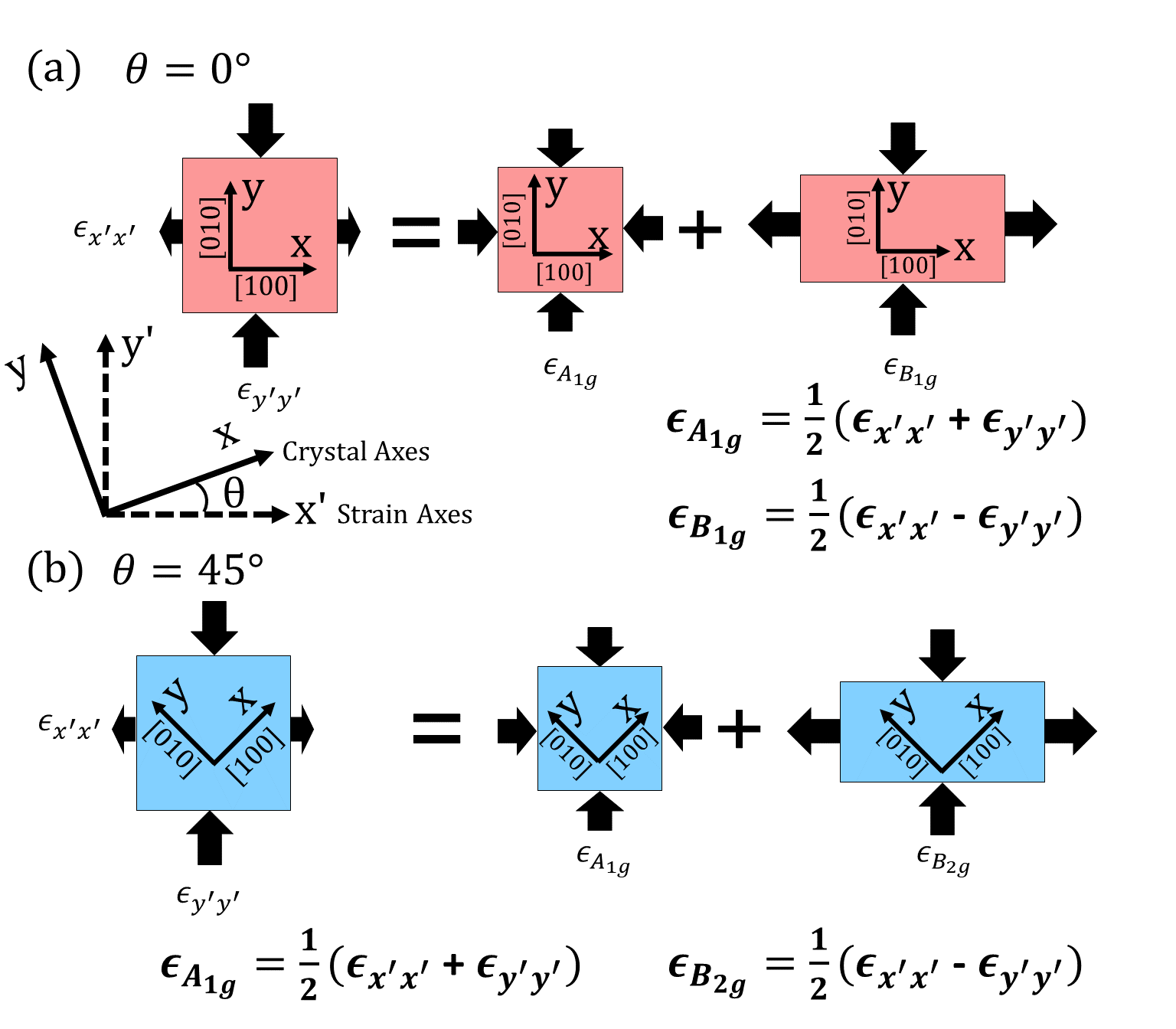}
\caption{\small \sl Schematic diagrams illustrating the symmetry decomposition of strains induced by the piezo stack. The unprimed coordinate system refers to the crystallographic axes and the primed coordinate system describes the normal strain frame, $\Theta$ is the angle between the two axes (inset). The normal (shearless) in-planes strains $\epsilon_{x'x'}$ and $\epsilon_{y'y'}$ are decomposed into components that are symmetric with respect to rotation about the $z$ axis ($\epsilon_{A_{1g}} = \frac{1}{2}(\epsilon_{x'x'} + \epsilon_{y'y'}$)) and antisymmetric with respect to rotation about the $z$-axis ($\epsilon_{B_{1g/2g}} = \frac{1}{2}(\epsilon_{x'x'} - \epsilon_{y'y'}$)). For the case shown in (a), where $\Theta=0^o$, such that the crystal axes are oriented along the normal strain frame, the sample experiences a normal antisymmetric ($B_{1g}$ symmetry) strain; $\epsilon_{B_{1g}} =\frac{1}{2}(\epsilon_{xx} - \epsilon_{yy}) = \frac{1}{2}(\epsilon_{x'x'} - \epsilon_{y'y'})$. For the case shown in (b), where $\Theta=45^o$, such that the crystal axes are rotated 45 degrees with respect to the normal strain frame, the sample experiences an antisymmetric shear strain ($B_{2g}$ symmetry); $\epsilon_{B_{2g}} =\epsilon_{xy} =\frac{1}{2}(\epsilon_{x'x'} - \epsilon_{y'y'})$. The representation of the normal and shear strains are the same in the strain frame. The ratio of the symmetric and antisymmetric strains is dictated by the in-plane Poisson ratio $\nu_P$ of the PZT stack ($\epsilon_{A_{1g}}=\frac{1-\nu_P}{1+\nu_P}\epsilon_{B_{1g/2g}}$).}

\label{fig:Intro}  
\end{center}  
\end{figure}

The Poisson ratio ($\nu_P$) of the PZT stack and the orientation of the crystallographic axes with respect to normal strain frame determines the symmetry of strain experienced by the crystal. This is illustrated in Fig. \ref{fig:Intro}. The Poisson ratio determines the ratio of strain along the $x'$ and $y'$ strain axes ($\epsilon_{y'y'}=-\nu_P\epsilon_{x'x'}$). A typical Poisson ratio for the PZT stacks used in these experiments is $\sim2.3$; therefore, the magnitude of strain along the $y'$ axis is larger than the magnitude of strain along the $x'$ axis. The strain can then be decomposed into two symmetry components: isotropic ($\epsilon_{A_{1g}}=\frac{1}{2}(\epsilon_{x'x'}+\epsilon_{y'y'})$) which is symmetric with respect to a $90^o$ rotation about the $z$ axis and antisymmetric ($\epsilon_{B_{1g/2g}}=\frac{1}{2}(\epsilon_{x'x'}-\epsilon_{y'y'})$) which is odd with respect to a $90^o$ rotation about the $z$-axis. The isotropic strain experienced by the crystal ($\epsilon_{A_{1g}}$) is independent of $\Theta$, the angle between the crystallographic axes and the normal strain frame; however, $\Theta$ determines the symmetry of the antisymmetric strain. For $\Theta=0^o$ the antisymmetric strain is normal ($B_{1g}$) (Fig. \ref{fig:Intro}(a)) and for $\Theta=45^o$ the antisymmetric strain is purely shear ($B_{2g}$) (Fig. \ref{fig:Intro}(b)).

\section{Strain Transmission}\label{sec:strain}

\begin{figure}[!hb]  
\begin{center}  
\includegraphics[width=4in]{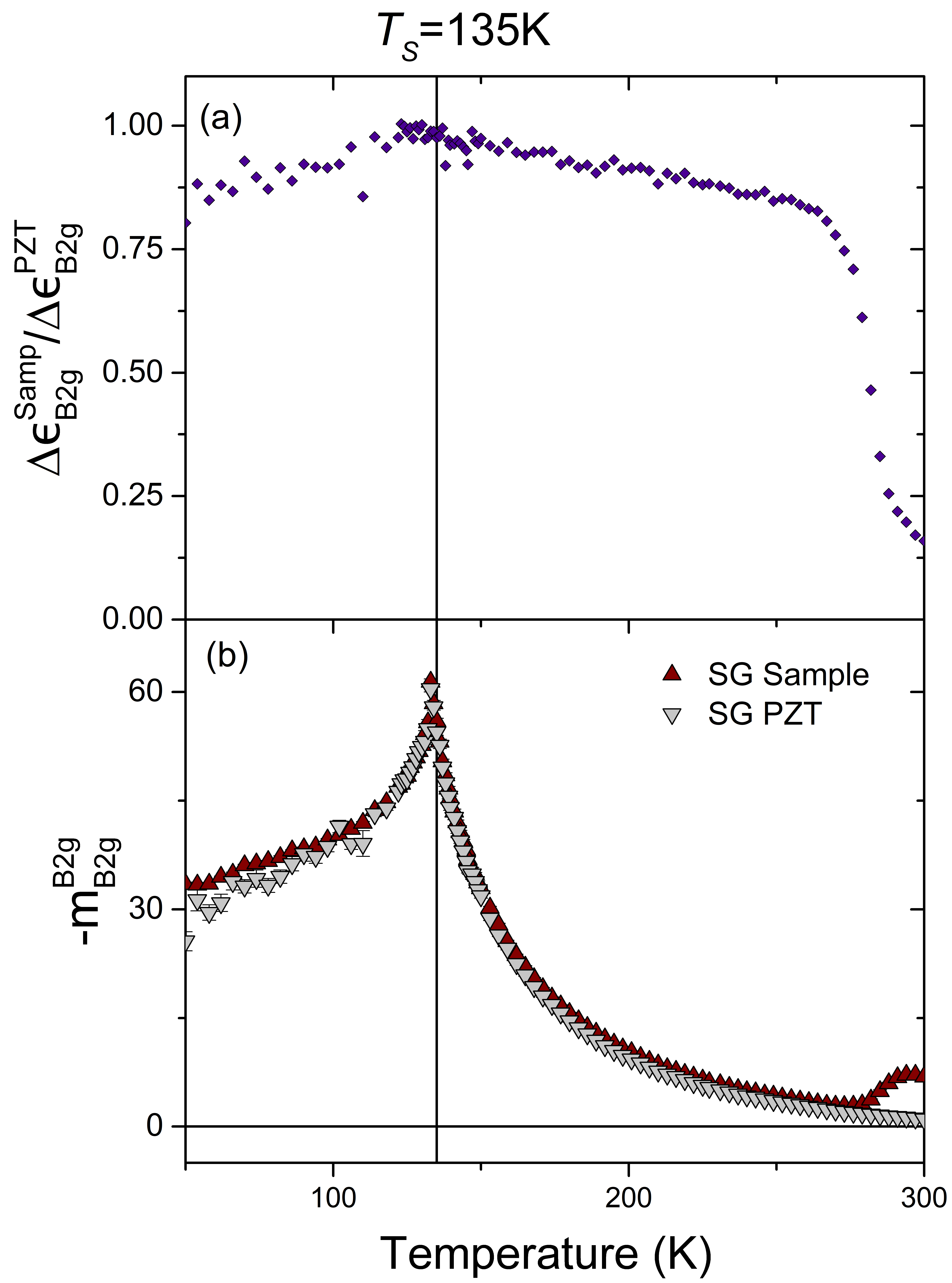}
\caption{\small \sl Strain transmission through a large (3140 $\mu m$ $\times$ 3330 $\mu m$ $\times$ 50 $\mu m$) BaFe$_2$As$_2$ crystal under $A_{1g}+B_{2g}$ symmetry strains. A strain gauge is glued on top of the sample ($\Delta\epsilon^{Samp}_{y'y'}$) and a second strain gauge is affixed directly to the back of the PZT stack ($\Delta\epsilon^{PZT}_{y'y'}$). For this particular test, $\Delta\epsilon_{x'x'}$ is estimated based on the measured Poisson ratio of the PZT, allowing for the estimation of the antisymmetric strains $\Delta\epsilon^{PZT}_{B_{2g}}$ and $\Delta\epsilon^{Samp}_{B_{2g}}$; $\frac{1}{2}(\Delta\epsilon^{PZT}_{x'x'}-\Delta\epsilon^{PZT}_{y'y'})$ and $\frac{1}{2}(\Delta\epsilon^{Samp}_{x'x'}-\Delta\epsilon^{Samp}_{y'y'})$ respectively. (a) The temperature dependence of the ratio of the range of antisymmetric strain experienced by the two strain gauges during fixed temperature voltage sweeps. Below 250K the strain transmission through the samples is $\geq 80\%$ and only has a weak temperature dependence. (b) The extracted $m^{B_{2g}}_{B_{2g}}$ elastoresistivity response calculated from both strain gauges. The two traces are in good agreement below 250K, indicating that the temperature dependence of the response is dominated by the intrinsic temperature dependence of the electronic sample properties over the temperature dependence of the strain transmission.}

\label{fig:SGComp}  
\end{center}  
\end{figure}

\begin{figure}[!b]  
\begin{center}  
\includegraphics[width=4in]{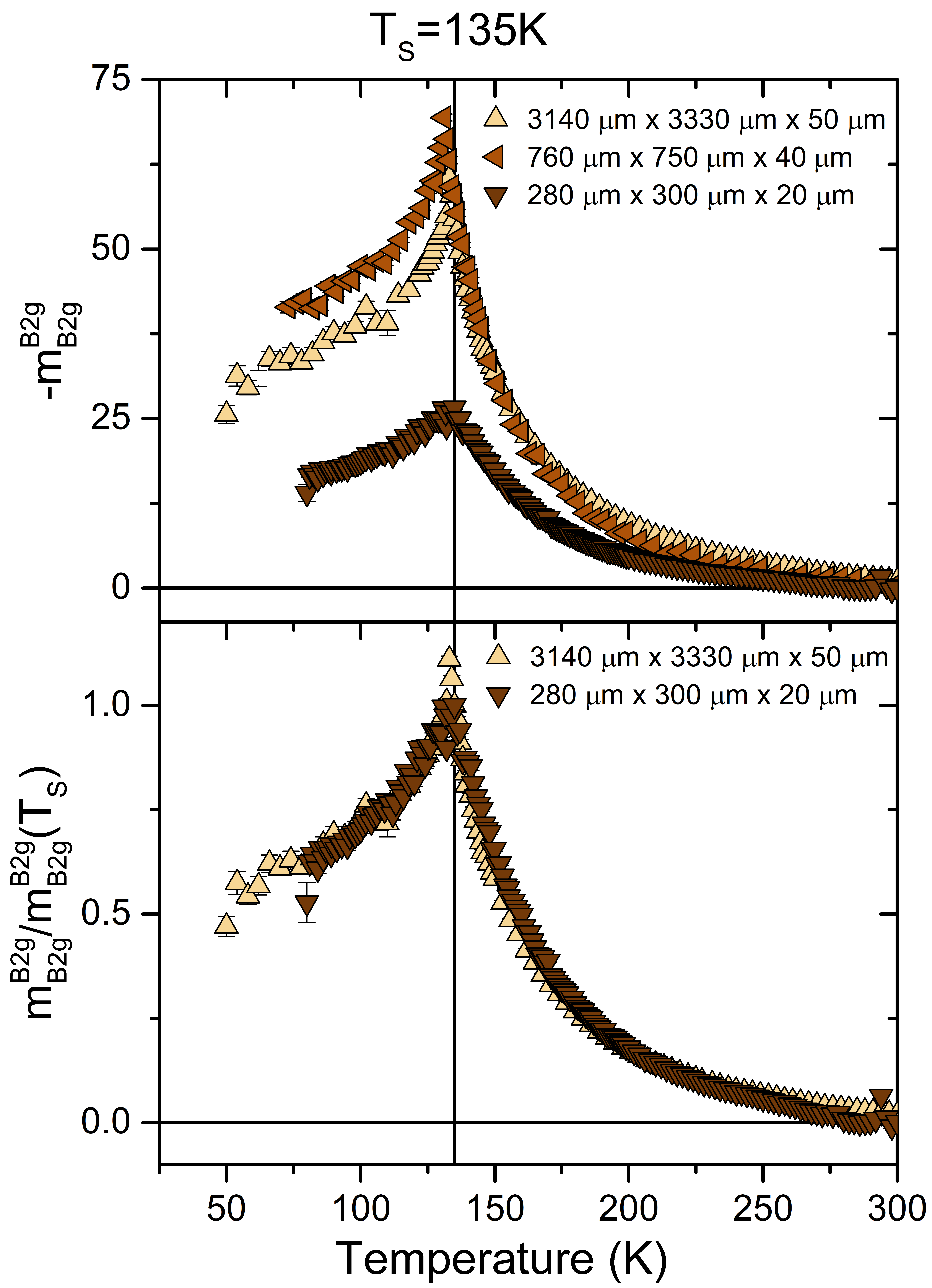}
\caption{\small \sl Comparison of sample size on the linear anisotropic elastoresistivity response, $m^{B_{2g}}_{B_{2g}}$ and strain transmission in BaFe$_2$As$_2$. Three sample sizes are studied: small (280 $\mu m$ $\times$ 300 $\mu m$ $\times$ 20 $\mu m$), medium (3760 $\mu m$ $\times$ 750 $\mu m$ $\times$ 40 $\mu m$), and large (3140 $\mu m$ $\times$ 3330 $\mu m$ $\times$ 50 $\mu m$). The strain at the surface of the large sample has been measured to be $\geq80\%$ below 250K (Fig. \ref{fig:SGComp}). For the data shown here the strain is measured by a strain gauge glued directly to the PZT stack and strain transmission is assumed to be $100\%$. The top plot (a) shows $m^{B_{2g}}_{B_{2g}}$ data for all three samples. The elastoresistivity responses of the medium and large samples have the same magnitude and temperature dependence suggesting they are in a regime of similar strain transmission ($\geq80\%$). The small sample has a significantly smaller response. This is attributed to imperfect strain transmission in the smallest sample, resulting in the overestimation of the strain experienced by the sample. The bottom plot (b) shows the normalized elastoresistivity response for the small and large sample. The two curves exhibit the same temperature dependence, indicating that imperfect strain transmission results in a temperature independent scaling of the response. } 

\label{fig:SizeComp}  
\end{center}  
\end{figure}

The strain transmission through the crystal will depend on geometric factors; for example, the thicker the crystal is compared with the in-plane dimensions, the more the strain will relax along the $z$-axis of the crystal. In order to quantify the strain transmission we compare a strain gauge mounted on top of a large undoped BaFe$_2$As$_2$ sample prepared as described in Sec. \ref{sec:method} and a strain gauge glued directly to the back of the PZT stack. For this experiment we measure the range of strain along the $y'$ direction, $\Delta\epsilon_{y'y'}^{Samp}$ and $\Delta\epsilon_{y'y'}^{PZT}$ for the sample mounted and PZT mounted strain gauges respectively, for fixed temperature voltage sweeps. The range of strain along the $x'$ direction is estimated based on the Poisson ratio of the PZT stack. From this we can calculate the range of antisymmetric strain experienced by the sample strain gauge ($\Delta\epsilon_{B_{2g}}^{Samp}=\frac{1}{2}(\Delta\epsilon^{Samp}_{x'x'}-\Delta\epsilon^{Samp}_{y'y'})$) and PZT strain gauge ($\Delta\epsilon_{B_{2g}}^{PZT}=\frac{1}{2}(\Delta\epsilon^{PZT}_{x'x'}-\Delta\epsilon^{PZT}_{y'y'})$). The temperature dependence of the ratio $\Delta\epsilon_{B_{2g}}^{Samp}/\Delta\epsilon_{B_{2g}}^{PZT}$ is shown in Fig. \ref{fig:SGComp}(a), a ratio of one implies perfect strain transmission through the sample. At 270K there is a sharp increase in strain transmission which we attribute to a freezing transition of the glue. Below 250K the strain transmission is $\geq80\%$ and has only a weak temperature dependence. This temperature dependence is small compared with the temperature dependence of the elastoresistance response which is demonstrated in Fig. \ref{fig:SGComp}(b) where $m^{B_{2g}}_{B_{2g}}$ is calculated twice, once using the measured strain of the strain gauge mounted on the sample ($\epsilon^{Samp}_{B_{2g}}$) and once using the measured strain of the strain gauge mounted on the PZT stack ($\epsilon^{PZT}_{B_{2g}}$). The two calculations are in good agreement below 250K.

The majority of samples are too small to accommodate a strain gauge on their surface. To quantify the strain transmission as a function of sample size, three undoped BaFe$_2$As$_2$ samples: small (280 $\mu m$ $\times$ 300 $\mu m$ $\times$ 20 $\mu m$), medium (760 $\mu m$ $\times$ 750 $\mu m$ $\times$ 40 $\mu m$), and large (3140 $\mu m$ $\times$ 3330 $\mu m$ $\times$ 50 $\mu m$, this sample is large enough to have a strain gauge on its surface and is the sample shown in Fig. \ref{fig:SGComp}) were measured. The extracted $m^{B_{2g}}_{B_{2g}}$ responses are shown in Fig. \ref{fig:SizeComp}(a). In these calculations of $m^{B_{2g}}_{B_{2g}}$, strain was measured by a strain gauge glued to the back of the PZT stack and the strain transmission was assumed to be 100$\%$. The large and medium samples have the same temperature dependence and magnitude of response, indicating that both samples have similar strain transmission ($\geq 80\%$). While the magnitude of the response of the small sample is significantly reduced, likely due to an overestimation of the strain experienced by the sample. This implies that for the small sample there is a strain gradient along the $z$ crystallographic axis and that the sample experiences an $E_g$ shear strain ($\epsilon_{x'z'}, \epsilon_{y'z'}$). By normalizing the $m_{B_{2g}}^{B_{2g}}$ response at the structural transition, $T_S=135K$, the temperature dependence of the small and large samples can be compared. This is shown in Fig. \ref{fig:SizeComp}(b). The two normalized responses are in good agreement below 250K, which demonstrates that imperfect strain transmission results in only a simple scaling of the magnitude of the elastoresistance response. In addition, it implies that the strain gradient and the $E_g$ shear strain have negligible effects on the in-plane elastoresistivity. 

This allows us to use the magnitude of the $m^{B_{2g}}_{B_{2g}}$ response as an approximate measure of strain transmission, with the assumption that samples with in-plane dimensions $\sim$750 $\mu m$ or larger have $\geq 80\%$ strain transmission. For samples oriented to experience $B_{1g}$ strain estimating the overall strain transmission is more challenging. Rough estimates are made based off of their relative size compared to samples that experience $B_{2g}$ strain. Table \ref{tab:strain} lists the sample dimensions and estimated strains for the Modified Montgomery Ba(Fe$_{0.975}$Co$_{0.025}$)$_2$As$_2$ samples used in the main text ($B_{2g}$ Sample 1 and $B_{1g}$ Sample 2) and this supplemental material.

\begin{table}[!h]
\begin{center}
\renewcommand{\arraystretch}{1.5} 
\begin{tabular}{c c c c}

\hline
\hline
\multicolumn{1}{c}{Orientation}  &  \multicolumn{1}{c}{Sample} & \multicolumn{1}{c}{Sample Dimensions ($\mu m$)}  & \multicolumn{1}{c}{Strain Transmission} \\
\hline
$B_{2g}$ & Sample 1 & $30 \times 730 \times 700$ & $\geq 80\%$ \\ 
$B_{2g}$ & Sample 2 & $15 \times 430 \times 430$ & $\geq 62\%$ \\ 
$B_{2g}$ & Sample 3 & $15 \times 400 \times 380$ & $\geq 53\%$ \\ 
$B_{1g}$ & Sample 1 & $15 \times 550 \times 500$ & $60\%-80\%$ \\ 
$B_{1g}$ & Sample 2 & $10 \times 540 \times 530$ & $60\%-80\%$ \\ 
 
\hline
\hline

\end{tabular}
\end{center}
\caption{\small \sl Sample dimensions and estimated strain transmission for the Ba(Fe$_{0.975}$Co$_{0.025}$)$_2$As$_2$ samples measured with the Modified Montgomery method used in the main paper and supplemental material. The samples used in the main paper are $B_{2g}$ Sample 1 and $B_{1g}$ Sample 2; all samples are shown in the supplemental material. The strain transmission for the samples that experience $B_{2g}$ symmetry strain are estimated from the magnitude of their $m_{B_{2g}}^{B_{2g}}$ elastoresistivity response with the assumption that samples with in-plane dimensions greater than or equal to $\sim750\mu m \times 750\mu m$ have $\geq 80\%$ strain transmission. The strain transmission for samples oriented to experience $B_{1g}$ symmetry strain is estimated by comparing relative sample size to the samples that experience $B_{2g}$ symmetry strain.} 
\label{tab:strain}
\end{table}

\section{Errors in Extracting the Linear and Quadratic Response from $\rho_{A_{1g}}$ arising from uncertainty in identifying the neutral strain point}

\begin{figure}[!h]  
\begin{center}  
\includegraphics[width=6in]{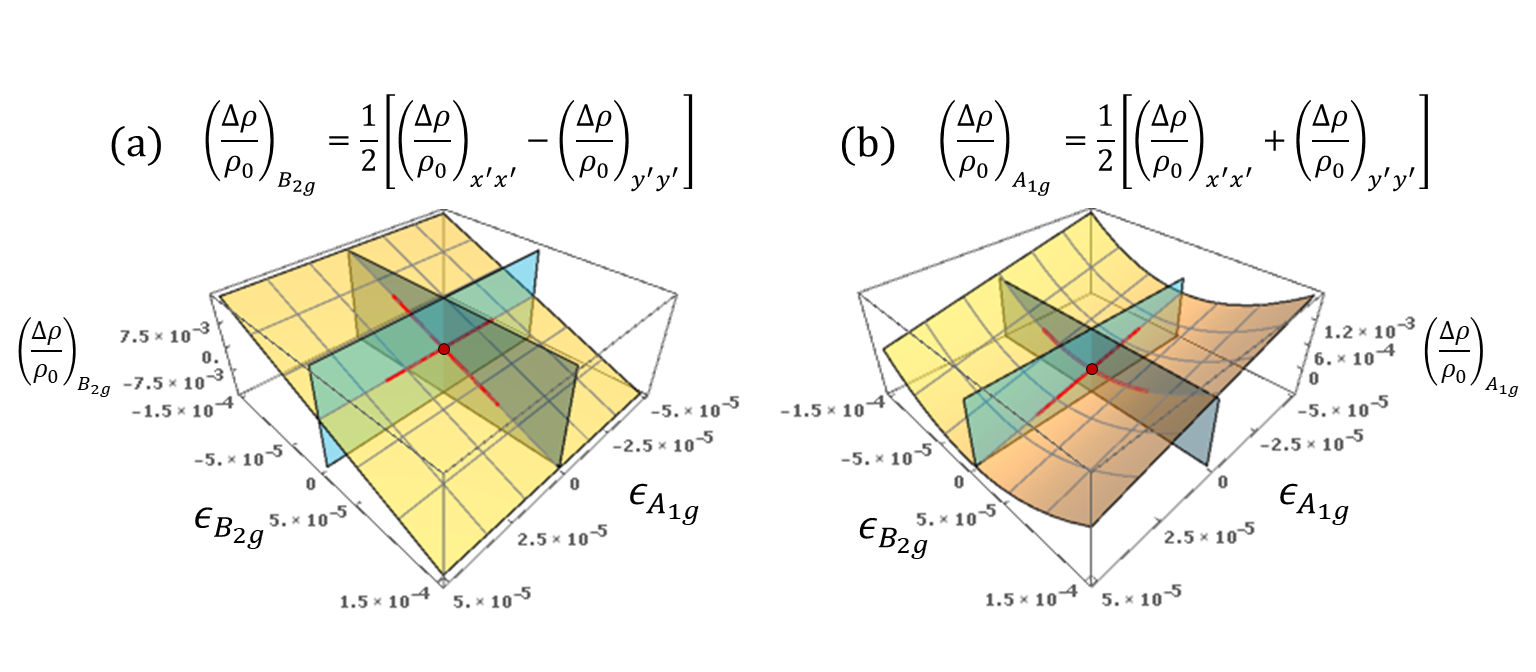}
\caption{\small \sl Schematic diagrams showing symmetry decomposed elastoresistivity responses to $A_{1g}$ and $B_{2g}$ strains. Vertical blue planes and red lines illustrate the response to strains with pure $A_{1g}$ and $B_{2g}$ symmetry strain. Red dots mark the neutral strain point ($\epsilon_{A_{1g}}=\epsilon_{B_{2g}}=0$). Panel (a) illustrates the antisymmetric elastoresistivity response $(\frac{\Delta\rho}{\rho_0})_{B_{2g}}$ as a function of $\epsilon_{A_{1g}}$ and $\epsilon_{B_{2g}}$ for a sample oriented with its crystallographic axes rotated 45 degrees from the normal strain frame (as shown Fig. \ref{fig:Intro}(b)). $A_{1g}$ symmetry strain cannot produce a response in the $B_{2g}$ symmetry channel, whereas $B_{2g}$ symmetry strain results in a linear response for the range of strains considered. The slope of the $B_{2g}$ symmetry resistivity response to $B_{2g}$ symmetry strain is linearly proportional to the nematic susceptibility in this symmetry channel. Panel (b) shows the symmetric elastoresistivity response $(\frac{\Delta\rho}{\rho_0})_{A_{1g}}$ for the same sample held under the same strain conditions. The $A_{1g}$ symmetry resistivity response to $A_{1g}$ symmetry strain is linear for the range of strain considered here. To linear order there is no response to $B_{2g}$ symmetry strain, but there is a quadratic response. For both panels the coefficients for the elastoresistance correspond to those measured for the title compound, Ba(Fe$_{0.975}$Co$_{0.025}$)$_2$As$_2$ at a temperature of 100 K.} 

\label{fig:SOM_Manifold}  
\end{center}  
\end{figure}

\begin{figure}[!h]  
\begin{center}  
\includegraphics[width=4in]{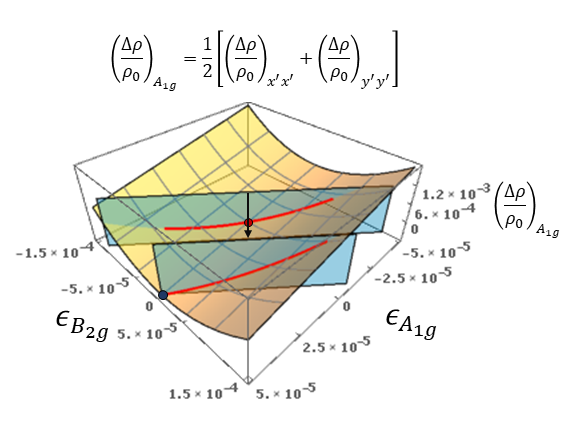}
\caption{\small \sl Schematic diagram of the isotropic resistivity response, $(\frac{\Delta\rho}{\rho_0})_{A_{1g}}$ to $A_{1g}$ and $B_{2g}$ symmetry strains. Vertical blue planes and red lines illustrate the response to representative voltage sweeps of the PZT stack. The ratio of $\epsilon_{A_{1g}}$ to $\epsilon_{B_{2g}}$ is determined by the in-plane Poisson ratio of the PZT stack ($\epsilon_{A_{1g}}=\frac{1-\nu_P}{1+\nu_P}\epsilon_{B_{2g}}$). One sweep is centered at $\epsilon_{B_{2g}}$=$\epsilon_{A_{1g}}=0$ (red dot). Due to thermal expansion differences between the glue, sample, and PZT stack and the volume contraction of the glue when it dries the zero strain point is often offset from zero volts applied to the PZT stack and the zero antisymmetric strain may not occur at zero isotropic strain. The second sweep shows the elastoresistivity response of a nonzero strain offset indicated by the black arrow, the neutral $\epsilon_{B_{2g}}$ strain point is marked by the blue dot. In both sweeps the minimum of the isotropic resistivity response is now offset from the zero antisymmetric strain point due to a contribution from the linear isotropic response. The coefficients for the elastoresistance correspond to those measured for Ba(Fe$_{0.975}$Co$_{0.025}$)$_2$As$_2$ at a temperature of 100 K.} 

\label{fig:SOM_PZTSweep}  
\end{center}  
\end{figure}

As shown in the main text, the main finding of the current work is that for the strain ranges we employ the elastoresistance of Ba(Fe$_{0.975}$Co$_{0.025}$)$_{2}$As$_2$ is linear with the exception of a large nonlinear $m_{A_{1g}}^{B_{2g}, B_{2g}}$ term. This is visualized in Fig. \ref{fig:SOM_Manifold} where we plot the symmetric and antisymmetric resistivity responses of Ba(Fe$_{0.975}$Co$_{0.025}$)$_2$As$_2$ to $A_{1g}$ and $B_{2g}$ symmetry strains. Elastoresistivity coefficients are taken from the measured compound at 100K. Verticle blue planes represent cuts of pure $A_{1g}$ or $B_{2g}$ symmetry strains. Red lines show the resistivity change along those cuts. The absolute neutral strain point ($\epsilon_{A_{1g}}=\epsilon_{B_{2g}}=0$) is marked by a red dot. In reality our strain sweeps may not be centered at $\epsilon_{A_{1g}}=\epsilon_{B_{2g}}=0$ due to differences in the thermal expansion of the PZT, sample, and the glue holding the sample in place and the volume contraction of the glue as it dries when the sample is attached to the PZT stack. The neutral $B_{2g}$ and $A_{1g}$ strain points may even be offset from each other. Since the PZT applies a fixed ratio of symmetric and antisymmetric strains if there is an offset in the neutral points at best we can tune through one neutral point at a time (i.e. $\epsilon_{A_{1g}}=0$ or $\epsilon_{B_{2g}}=0$). This is shown in Fig. \ref{fig:SOM_PZTSweep}. Blue planes now sweep along the fixed ratio of antisymmetric to symmetric strain for a typical sweep of the PZT stack at 100K. Two sweeps are shown, one that crosses the absolute neutral point (red dot, $\epsilon_{A_{1g}}=\epsilon_{B_{2g}}=0$) and one that sweeps through the antisymmetric strain neutral point at finite symmetric strain (blue dot, $\epsilon_{B_{2g}}=0$, $\epsilon_{A_{1g}}=5\times10^{-5}$).

We can identify the neutral antisymmetric strain point above the tetragonal to orthorhombic structural transition (98K) because, for a tetragonal material at neutral antisymmetric strain point, $\rho_{xx}=\rho_{yy}$. The Modified Montgomery method is well suited to identify the antisymmetric strain neutral point since it simultaneously measures $\rho_{x'x'}$ and $\rho_{y'y'}$ under identical strain conditions in a single sample. This is one advantage of the Modified Montgomery method over the previously used differential technique \cite{kuo2013}. It is more challenging to identify the symmetric strain neutral point and it is not done in this work. Below is a detailed calculation of the effects of the misidentification of strain offsets on the calculated elastoresistivity tensor components. The main results are that, for this material, the correct identification of the antisymmetric neutral point is required to accurately estimate $m_{A_{1g}}^{A_{1g}}$ for samples that experience $A_{1g}$ and $B_{2g}$ symmetry strain, however neither $m_{B_{1g/2g}}^{B_{1g/2g}}$ nor $m^{B_{1g/2g},B_{1g/2g} }_{A_{1g}}$ are dependent on the identification of the neutral point. All results are robust to the determination of the symmetric strain neutral point.

Lets start with the simple case of the linear antisymmetric response, assuming no offset between the neutral $A_{1g}$ and $B_{1g/2g}$ strain points and that $\epsilon_{A_{1g}}$, $\epsilon_{B_{1g/2g}}$, and $\epsilon_{x'x'}$ are all measured relative to the neutral point where $\epsilon_{A_{1g}} = \epsilon_{B_{1g/2g}} = \epsilon_{x'x'} = 0$. Then the change in antisymmetric resistivity to $\epsilon_{x'x'}$ is described by, 

\begin{equation}
\begin{split}
(\Delta\rho/\rho_0)_{B_{1g/2g}}&=m^{B_{1g/2g}}_{B_{1g/2g}}\epsilon_{B_{1g/2g}}\\&=m^{B_{1g/2g}}_{B_{1g/2g}}(\frac{1+\nu_P}{2})\epsilon_{x'x'}
\end{split}
\end{equation}
If the neutral strain point is misidentified by an amount $\Delta\epsilon_{x'x'}$ such that $\epsilon_{x'x'}^{true}=\epsilon_{x'x'}^{measured} + \Delta\epsilon_{x'x'}$ then there will be an offset in both the symmetric and antisymmetric neutral points (i.e. $\Delta\epsilon_{A_{1g}}=\frac{1-\nu_P}{2}\Delta\epsilon_{x'x'}$ and $\Delta\epsilon_{B_{2g}}=\frac{1+\nu_P}{2}\Delta\epsilon_{x'x'}$). Then the antisymmetric response becomes,

\begin{equation}
(\Delta\rho/\rho_0)_{B_{1g/2g}}=m^{B_{1g/2g}}_{B_{1g/2g}}(\frac{1+\nu_P}{2})(\epsilon^{measured}_{x'x'}+\Delta\epsilon_{x'x'})
\end{equation}
The linear antisymmetric elastoresistivity coefficient is extracted from the slope of the linear fit of $(\Delta\rho/\rho_0)_{B_{1g/2g}}$ vs $\frac{1+\nu_P}{2}\epsilon^{measured}_{x'x'}$ ($\epsilon_{B_{1g/2g}}^{measured}$). In this case the extracted slope is the true elastoresistivity coefficient, $m^{B_{1g/2g}}_{{B_{1g/2g}}}$, independent of the error in the identification of the strain neutral point $\Delta\epsilon_{x'x'}$. 

For the isotropic resistivity response, we again start by assuming no offset between the neutral symmetric and antisymmetric strain points and that all strains are measured relative to the neutral point where $\epsilon_{A_{1g}} = \epsilon_{B_{1g/2g}} = \epsilon_{x'x'} = 0$. For simplicity we will perform these calculations for a sample that experiences $A_{1g}$ and $B_{2g}$ symmetry strain (the same calculation can be done for a sample that experiences $A_{1g}$ and $B_{1g}$ symmetry strains by simply replacing all references to $B_{2g}$ with $B_{1g})$. The isotropic resistivity response is then described by, 

\begin{equation}
\begin{split}
(\Delta\rho/\rho_0)_{A_{1g}} &= m^{A_{1g}}_{A_{1g}}\epsilon_{A_{1g}} + m_{A_{1g}}^{A_{1g}, A_{1g}}\epsilon^2_{A_{1g}} + m_{A_{1g}}^{B_{2g}, B_{2g}}\epsilon^2_{B_{2g}}\\
&= m^{A_{1g}}_{A_{1g}}\epsilon_{x'x'}(\frac{1-\nu_P}{2}) + m_{A_{1g}}^{A_{1g}, A_{1g}}\epsilon^2_{x'x'}(\frac{1-\nu_P}{2})^2 + m_{A_{1g}}^{B_{2g}, B_{2g}}\epsilon^2_{x'x'}(\frac{1+\nu_P}{2})^2
\end{split}
\end{equation}
Now we introduce a misidentification of the neutral strain point by an amount $\Delta\epsilon_{x'x'}$ ($\epsilon_{x'x'}^{true}=\epsilon_{x'x'}^{measured} + \Delta\epsilon_{x'x'}$). The isotropic resistivity response then becomes, 


\begin{equation}
\begin{split}
(\Delta\rho/\rho_0)_{A_{1g}} =&((\frac{1-\nu_P}{1+\nu_P})^2m^{A_{1g}, A_{1g}}_{A_{1g}} + m^{B_{2g}, B_{2g}}_{A_{1g}})[\frac{1+\nu_P}{2}\epsilon^{measured}_{x'x'}]^2\\&+(m^{A_{1g}}_{A_{1g}} + (1-\nu_P)m^{A_{1g},A_{1g}}_{A_{1g}}\Delta\epsilon_{x'x'}+\frac{(1+\nu_P)^2}{1-\nu_P}m^{B_{2g}, B_{2g}}_{A_{1g}}\Delta\epsilon_{x'x'})\frac{1-\nu_P}{2}\epsilon^{measured}_{x'x'}\\&+\frac{1-\nu_P}{2}m^{A_{1g}}_{A_{1g}}\Delta\epsilon_{x'x'} + (\frac{1-\nu_P}{2})^2m^{A_{1g}, A_{1g}}_{A_{1g}}[\Delta\epsilon_{x'x'}]^2 + (\frac{1+\nu_P}{2})^2m^{B_{2g},B_{2g}}_{A_{1g}}[\Delta\epsilon_{x'x'}]^2 
\end{split}
\end{equation}
For this material the only non-negligible quadratic response is $m^{B_{2g}, B_{2g}}_{A_{1g}}$ ($m^{B_{1g}, B_{1g}}_{A_{1g}}\approx m^{A_{1g}, A_{1g}}_{A_{1g}}\approx0$). This further simplifies the equation,

\begin{equation}
\begin{split}
(\Delta\rho/\rho_0)_{A_{1g}} =&m^{B_{2g}, B_{2g}}_{A_{1g}}[\frac{1+\nu_P}{2}\epsilon^{measured}_{x'x'}]^2+(m^{A_{1g}}_{A_{1g}} +\frac{(1+\nu_P)^2}{1-\nu_P}m^{B_{2g}, B_{2g}}_{A_{1g}}\Delta\epsilon_{x'x'})\frac{1-\nu_P}{2}\epsilon^{measured}_{x'x'}\\&+\frac{1-\nu_P}{2}m^{A_{1g}}_{A_{1g}}\Delta\epsilon_{x'x'} + (\frac{1+\nu_P}{2})^2m^{B_{2g},B_{2g}}_{A_{1g}}[\Delta\epsilon_{x'x'}]^2 
\end{split}
\end{equation}
Fits to the linear $(\frac{\Delta\rho}{\rho_0})_{A_{1g}}$ response vs $\frac{1-\nu_P}{2}\epsilon_{x'x'}^{measured}$ ($\epsilon^{measured}_{A_{1g}}$) incorrectly identify the slope, the effective measured $m^{A_{1g}}_{A_{1g}}$, as $m^{A_{1g}}_{A_{1g}} + \frac{(1+\nu_P)^2}{1-\nu_P}m_{A_{1g}}^{B_{2g}, B_{2g}}\Delta\epsilon_{x'x'} = m^{A_{1g}}_{A_{1g}} + 2(\frac{1+\nu_P}{1-\nu_P})m_{A_{1g}}^{B_{2g}, B_{2g}}\Delta\epsilon_{B_{2g}}$, so to accurately measure this quantity the neutral $B_{2g}$ strain point must be correctly identified. If a similar procedure is followed for a sample experiencing $A_{1g}$ and $B_{1g}$ symmetry strain there is no error introduced to the measured $m^{A_{1g}}_{A_{1g}}$ for misidentification of the neutral strain point or for offsets between the $B_{1g}$ and $A_{1g}$ neutral points since there is no contribution from the quadratic response. Thus estimates of $m^{A_{1g}}_{A_{1g}}$ extracted from samples that experience $A_{1g}$ and $B_{1g}$ symmetry strains are robust. Fits to the quadratic $(\frac{\Delta\rho}{\rho_0})_{A_{1g}}$ response vs $[\frac{1+\nu_P}{2}\epsilon_{x'x'}^{measured}]^2$ ($[\epsilon^{measured}_{B_{2g}}]^2$) correctly extract the quadratic coefficient, $m^{B_{2g}, B_{2g}}_{A_{1g}}$, independent of the neutral strain.  

Two experimental observations confirm that we can correctly identify the neutral $B_{2g}$ strain point (blue dot in Fig. \ref{fig:SOM_PZTSweep}). First the estimates of $m^{A_{1g}}_{A_{1g}}$ (shown in Fig. 3(b) of the main text) are the same for crystals oriented such that they exhibit $A_{1g}+B_{2g}$ and $A_{1g}+B_{1g}$ strains. Secondly, misidentification of the $B_{2g}$ neutral point would admix some amount of $m^{B_{2g}, B_{2g}}_{A_{1g}}$ into the nominal measurement of $m^{A_{1g}}_{A_{1g}}$, which would introduce a strong temperature dependence-- this is not observed. 

\begin{figure}[!h]  
\begin{center}  
\includegraphics[width=4in]{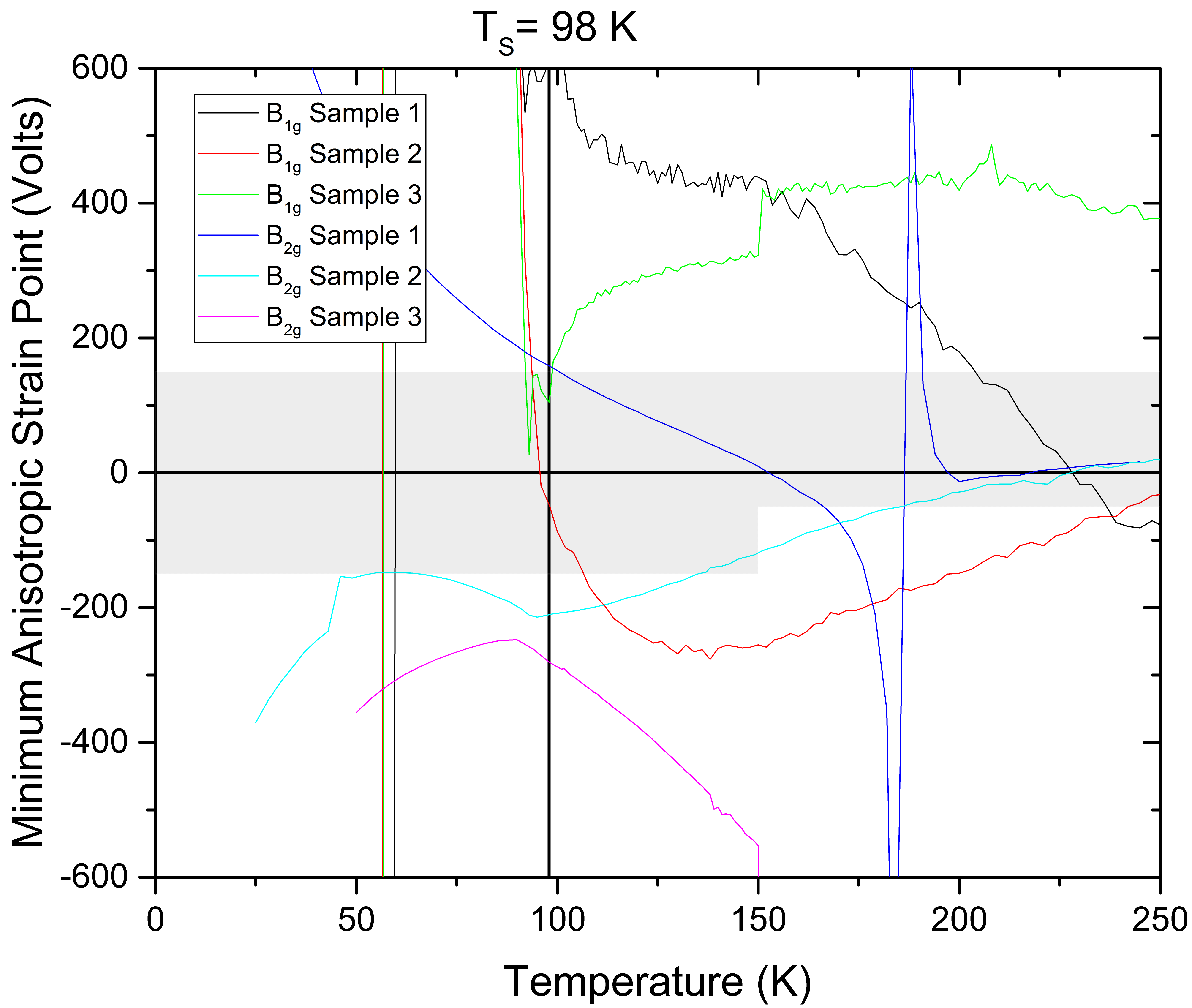}
\caption{\small \sl The temperature dependence of the zero anisotropic strain point as a function of voltage on the PZT stack. The shaded gray region shows the experimentally accessible voltage range, the zero anisotropic strain point is estimated from quadratically (linearly) fitting the response close to (far from) the accessible strain range. There is no common trend in the temperature dependence of the anisotropic neutral point, indicating that this effect is not produced solely by differential thermal contraction between the PZT and sample but that the epoxy plays a significant role.} 

\label{fig:SOM_Zero_Strain}  
\end{center}  
\end{figure}

Finally, the temperature dependence of the antisymmetric strain neutral point (as a function of voltage applied to the PZT stack) is plotted in Fig. \ref{fig:SOM_Zero_Strain} for six Ba(Fe$_{0.975}$Co$_{0.025}$)$_{2}$As$_2$ samples. There does not appear to be a common trend in the evolution of the neutral point as a function of temperature. This demonstrates that this effect is not solely due to differential thermal contractions of the sample on PZT, implying that the epoxy plays a significant role in determine the ``zero volts" strain experienced by the sample. 

\section{Temperature dependence of $m^{B_{1g}}_{B_{1g}}$ }

\begin{figure}[!h]  
\begin{center}  
\includegraphics[width=4in]{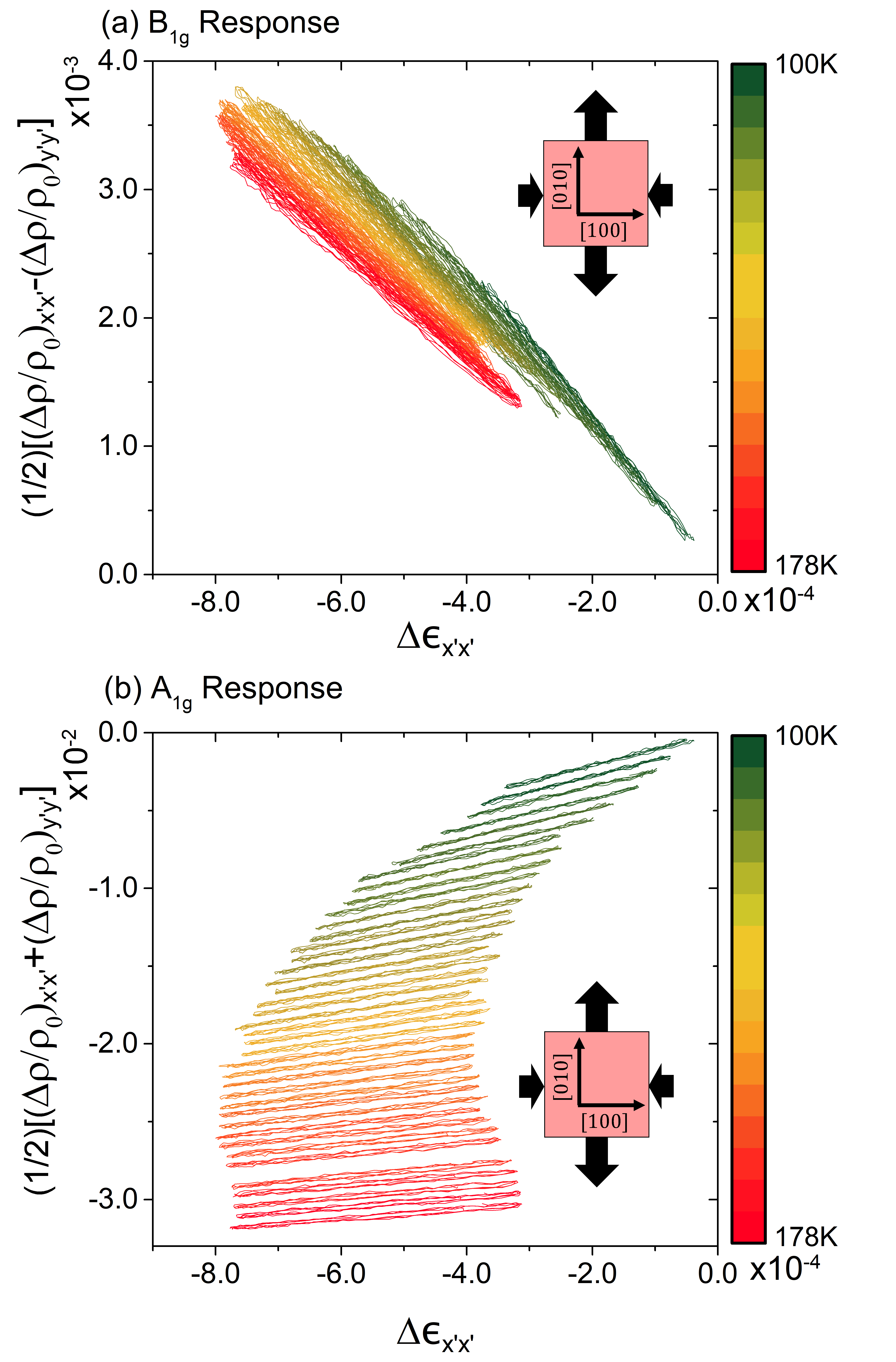}
\caption{\small \sl Temperature dependence of (a) the antisymmetric ($B_{1g}$) elastoresistivity response, and (b) the isotropic ($A_{1g}$) elastoresistivity response, of a single crystal of $Ba(Fe_{0.975}Co_{0.025})_{2}As_2$ oriented with the crystal axes normal to the strain frame (pink schematic insets). The antisymmetric and isotropic response is always linear with a weak temperature dependence. For clarity, data showing the isotropic response are offset by $-7.5*10^{-4}$ from the 100 K sweep. The data showing the antisymmetric response are not offset. Data shown in the upper panel are used to extract $m^{B_{1g}}_{B_{1g}}$ (shown in Fig. 3(a) of the main text). Data shown in the lower panel are used to extract $m^{A_{1g}}_{A_{1g}}$ and $(\frac{1-\nu_P}{1+\nu_P})^2m_{A_{1g}}^{A_{1g}, A_{1g}} + m_{A_{1g}}^{B_{1g}, B_{1g}}$ (shown in main text Fig. 3(b) and Fig. 3(c) respectively). The trace at 168K is not shown due to unusually large noise in that trace.}

\label{fig:SOM_Waterfall}  
\end{center}  
\end{figure}

The elastoresistance response of Ba(Fe$_{0.975}$Co$_{0.025}$)$_{2}$As$_2$ to $B_{1g}$ symmetry strain has a much weaker temperature dependence than the response to $B_{2g}$ symmetry strain. Figure \ref{fig:SOM_Waterfall} shows the anisotropic and isotropic resistivity response from 100 K to 179 K for $B_{1g}$ Sample 2. Both resistivity responses are linear for all temperatures. The zero anisotropic strain for this sample is farther from the accessible strain range then the sample experiencing $B_{2g}$ strain in the main text, however a quadratic response within one order of magnitude would still be clearly resolvable. No evidence of a quadratic response was seen for any of the three samples measured under $B_{1g}$ symmetry strains. As demonstrated in the main text and shown in main text Fig. 3(b), the linear response to $A_{1g}$ symmetry strain $m^{A_{1g}}_{A_{1g}}$ is identical within our experimental resolution for crystals oriented such that they experience $B_{1g} + A_{1g}$ strain or $B_{2g} + A_{1g}$ strain, as should be the case by symmetry.

\section{Multiple measurements of $m^{B_{2g}}_{B_{2g}}$}

\begin{figure}[!h]  
\begin{center}  
\includegraphics[width=4in]{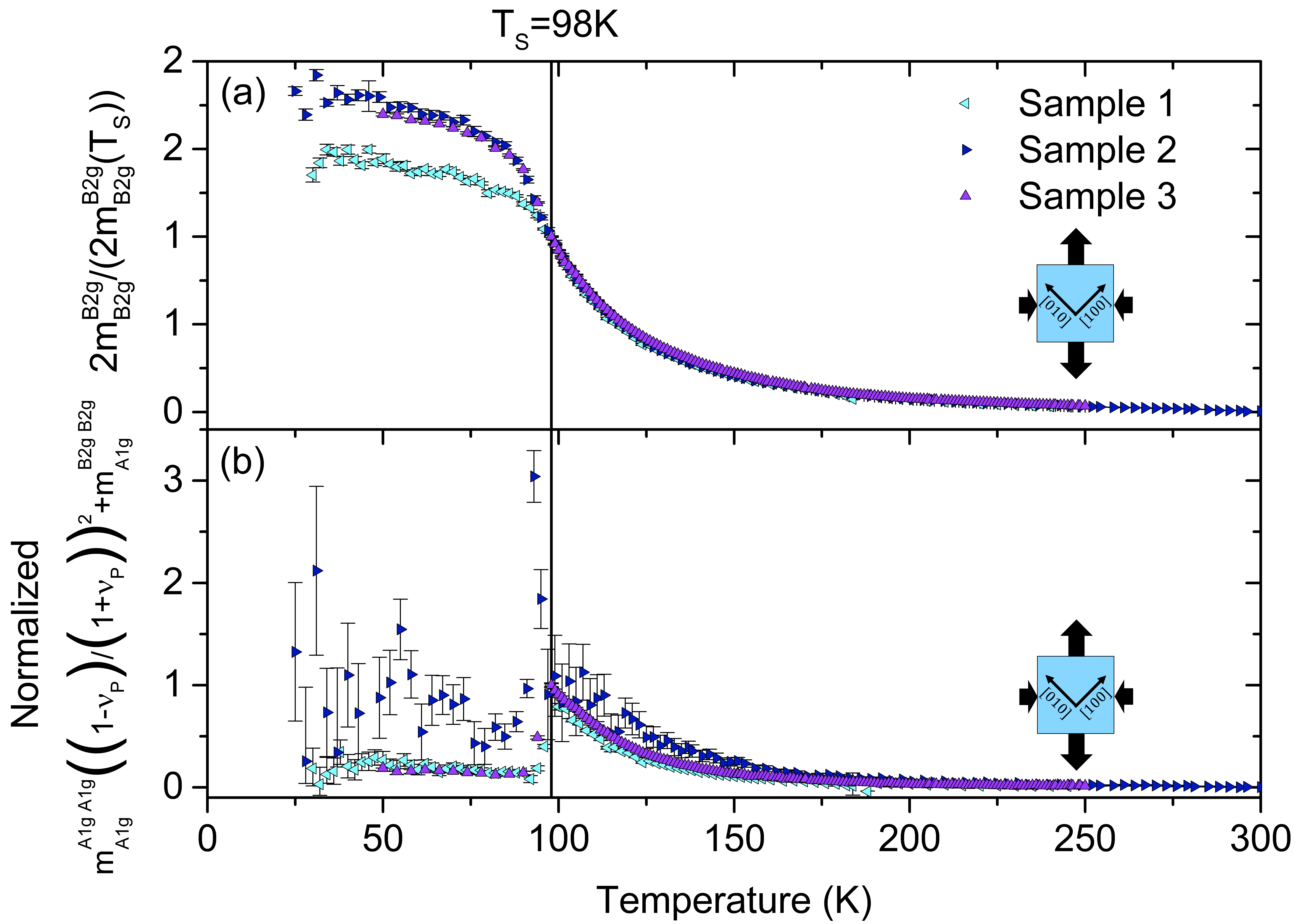}
\caption{\small \sl Comparison of the elastoresistance response of three Ba(Fe$_{0.975}$Co$_{0.025}$)$_{2}$As$_2$ samples under $A_{1g}$ and $B_{2g}$ symmetry strains and nominally identical experimental conditions. The responses are normalized at the structural transition to account for differences in strain transmission due to sample size and aspect ratio effects. Upper panel (a) shows the linear response to anisotropic strain, $m^{B_{2g}}_{B_{2g}}$. Lower panel (b) shows the extracted quadratic coefficient, dominated from contributions of $m_{A_{1g}}^{B_{2g}, B_{2g}}$. Data for all three samples are in good agreement. The origin of the large uncertainty (error bars) for sample 2 in panel (b) is discussed in the text. Sample 1 is used in the main paper.}

\label{fig:ER_B2g}  
\end{center}  
\end{figure}

The temperature dependence of the elastoresistance response is very reproducible from sample to sample. Fig. \ref{fig:ER_B2g} shows the temperature dependence of the normalized linear ($m^{B_{2g}}_{B_{2g}}$) and quadratic $((\frac{1-\nu_P}{1+\nu_P})^2m_{A_{1g}}^{A_{1g}, A_{1g}} + m_{A_{1g}}^{B_{2g}, B_{2g}})$ elastoresistance response for three samples oriented with the crystal axes 45 degrees with respect to the strain axes. Variations of the magnitude of the response between samples are dominated by sample geometry and incomplete strain transmission. This is discussed in detail in Sec. \ref{sec:strain}. 

There was an error in setting up the experiment for sample 2. The measurement was performed with an averaging time constant comparable to the sampling time between data points, this introduced a small hysteresis in the resistivity vs strain response. This has minimal impact in fitting the linear response, but introduces a systematic deviation when fitting the quadratic response. The combination of hysteresis and this particular sample being far from the neutral anisotropic strain point introduces large errors when fitting the quadratic response. These data are included in Fig. \ref{fig:ER_B2g} since they further corroborate our findings, though the uncertainty in each data point is larger than for samples 1 and 3. 

\section{Fitting $m^{B_{2g}}_{B_{2g}}$ and $m_{A_{1g}}^{B_{2g}, B_{2g}}$}

\begin{table}[h]
\begin{center}
\renewcommand{\arraystretch}{1.5} 
\begin{tabular}{m{30mm} m{25mm} m{25mm} m{27mm} m{12mm}}

\hline
\hline
\multicolumn{1}{c}{Function}  & \multicolumn{1}{c}{a}  & \multicolumn{1}{c}{b} & \multicolumn{1}{c}{c} & \multicolumn{1}{c}{$R^2_{Adj}$}\\
\hline

 $\frac{a}{(T-\Theta)^2} + \frac{b}{T-\Theta} + c$  & $(2.0 \pm 0.2) \times 10^7$ & $(7.4 \pm 1) 
\times 10^5$ & $(-6.7 \pm 0.9) \times 10^3$ & $0.9965$ \\ 
$\frac{a}{(T-\Theta)^2} + c$   & $(3.80 \pm 0.06) \times 10^7$ & \multicolumn{1}{c}{-} & $(0.8 \pm 3) \times 10^2$ & $0.9909$ \\ 
$\frac{b}{T-\Theta} + c$ & \multicolumn{1}{c}{-} & $(1.55 \pm 0.03) \times 10^6$  & $(1.39 \pm 0.05) \times 10^4$ & $0.9896$\\

\hline
\hline

\end{tabular}
\end{center}
\caption{\small \sl Extracted fit parameters and goodness of fit for three fitting methods of the weighted nonlinear isotropic resistivity response to antisymmetric strain, $(\frac{1-\nu_P}{1+\nu_P})^2m_{A_{1g}}^{A_{1g}, A_{1g}} + m_{A_{1g}}^{B_{2g}, B_{2g}}$, between 104K to 181K for Ba(Fe$_{0.975}$Co$_{0.025}$)$_{2}$A$_2$ under $A_{1g}$ and $B_{2g}$ symmetry strains. The Weiss temperature, $\Theta$, is fixed at 75.8K from the Curie-Weiss fit of the linear ansitropic response $m_{B_{2g}}^{B_{2g}}$ and is not a fit parameter. The temperature dependence of the fits is motivated by symmetry and the temperature dependence of the nematic susceptibility, see text for details. The best fit of the data is the functional form $\frac{a}{(T-\Theta)^2} + \frac{b}{T-\Theta} + c$.} 
\label{table:quad}
\end{table}

\begin{figure}[t]  
\begin{center}  
\includegraphics[width=3in]{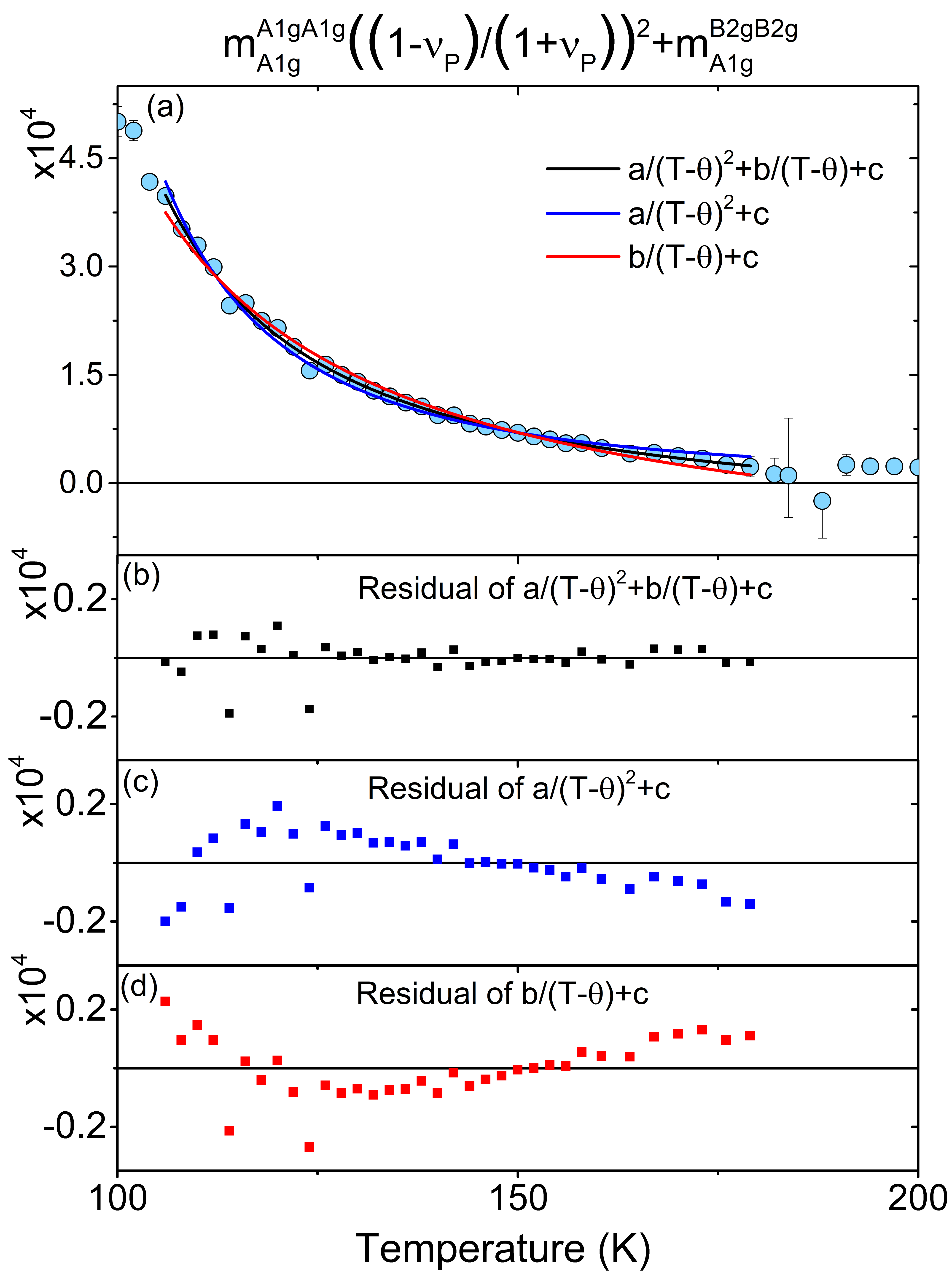}
\caption{\small \sl Fits of the extracted weighted quadratic response of the isotropic resistivity, $(\frac{1-\nu_P}{1+\nu_P})^2m_{A_{1g}}^{A_{1g}, A_{1g}} + m_{A_{1g}}^{B_{2g}, B_{2g}}$, for Ba(Fe$_{0.975}$Co$_{0.025}$)$_{2}$A$_2$ under $A_{1g}$ and $B_{2g}$ symmetry strains. The elastoresistivity response is fit between 104K to 181K and the Weiss temperature, $\Theta$, is fixed at 75.8K from the Curie-Weiss fit of the linear anisotropic response. The temperature dependence of the three fit equations is motivated by symmetry, see text for details. The top plot (a) shows an overlay of the three fits (solid lines) on the experimentally observed response (blue circles). The residuals for the three fits are shown in the lower plots (b)-(d). The residual for the fit of the functional form  $\frac{a}{(T-\Theta)^2} + \frac{b}{T-\Theta} + c$ (black dots) is shown in plot (b). The residual is flat as function of temperature and centered around zero. The residuals of the $\frac{a}{(T-\Theta)^2} + c$ (blue dots) and $\frac{b}{T-\Theta} + c$ (red dots) fits are shown in plots (c) and (d) respectively. Both have a clear temperature dependence, indicating neither fully captures the temperature dependence of the response. 
} 

\label{fig:SOM_Quad}  
\end{center}  
\end{figure}

The linear antisymmetric response to $B_{2g}$ symmetry strain, $m^{B_{2g}}_{B_{2g}}$, is extracted from a first order fit of $(\frac{\Delta\rho}{\rho_0})_{B_{2g}}$ versus $\epsilon_{B_{2g}}$. The temperature dependence of $m^{B_{2g}}_{B_{2g}}$ can then be fit to a Curie-Weiss functional form, $m^{B_{2g}}_{B_{2g}} = \frac{\lambda}{a_0}(\frac{1}{T-\Theta}) + m^{B_{2g}}_{B_{2g}, 0}$. The antisymmetric response deviates from a true Curie-Weiss behavior at high temperatures where the epoxy softens and at low temperatures due to the structural transition and disorder. The optimal temperature range to extract the best Curie-Weiss fit is chosen following the procedure outline in reference \cite{Kuo2016}, except that the reduced $\chi^2$ error was minimized as opposed to the standard deviation. For the sample shown in the main text the best fit temperature range was found to be 104K-181K. Extracted fit parameters are shown in table \ref{table:fits}.

The nonlinear symmetric response to antisymmetric $B_{2g}$ strain, $m_{A_{1g}}^{B_{2g}, B_{2g}}$, was extracted from the quadratic coefficient of a second order fit of  $(\frac{\Delta\rho}{\rho_0})_{A_{1g}}$ versus $\epsilon_{B_{2g}}$. As described in the main text, by symmetry the temperature dependence of $m_{A_{1g}}^{B_{2g}, B_{2g}}$ is allowed to include the terms $\frac{a}{(T-\Theta)^2} + \frac{b}{T-\Theta} + c$. For reasons beyond simple symmetry arguments, one of these terms may not contribute and a full expression would over fit the data. Fits for three combinations of terms ($\frac{a}{(T-\Theta)^2} + \frac{b}{T-\Theta} + c$, $\frac{a}{(T-\Theta)^2} + c$, and $\frac{b}{T-\Theta} + c$) are shown in Fig. \ref{fig:SOM_Quad}. The critical temperature, $\Theta$, is fixed to be 75.8K from the Curie-Weiss fit of $m^{B_{2g}}_{B_{2g}}$. The fitted coefficients are shown in table \ref{table:quad}. To capture the effects of over fitting we look at the $R^2_{adj}$ goodness of fit. The full form, $\frac{a}{(T-\Theta)^2} + \frac{b}{T-\Theta} + c$, best fits the data with an $R^2_{adj}$ of 0.9965. The residuals of $\frac{a}{(T-\Theta)^2} + c$ and the Curie-Weiss ($\frac{b}{T-\Theta} + c$) fits show clear systematic trends as a function of temperature, shown in Fig. \ref{fig:SOM_Quad}(c) and Fig. \ref{fig:SOM_Quad}(d) respectively. This indicates that the poorer $R^2_{adj}$ arises because these functional forms do not correctly describe the temperature dependence. In contrast the residuals for the fit to the full form ($\frac{a}{(T-\Theta)^2} + \frac{b}{T-\Theta} + c$) do not exhibit any systematic trends, suggesting that the data is well described by this functional form.  Fit parameters are listed in table \ref{table:quad}. We can also compare the magnitude of the contributions from individual terms in the fit. Close to the Weiss temperature (i.e. as $T-\Theta\rightarrow 0$), we expect the quadratic term ($\frac{a}{(T-\Theta)^2}$) to dominate; however, at high temperatures the Curie-Weiss term ($\frac{b}{T-\Theta}$) is largest. The cross over point is roughly 23K above the Weiss temperature, so for the accessible range of temperatures considered here (i.e. above $T_s=98$K) the Curie-Weiss term is equal in magnitude or larger than the $\frac{a}{(T-\Theta)^2}$ component.

\section{Comparison with the $45^o$ elastoresistance measurement technique}

\begin{figure}[!hb]  
\begin{center}  
\includegraphics[width=3in]{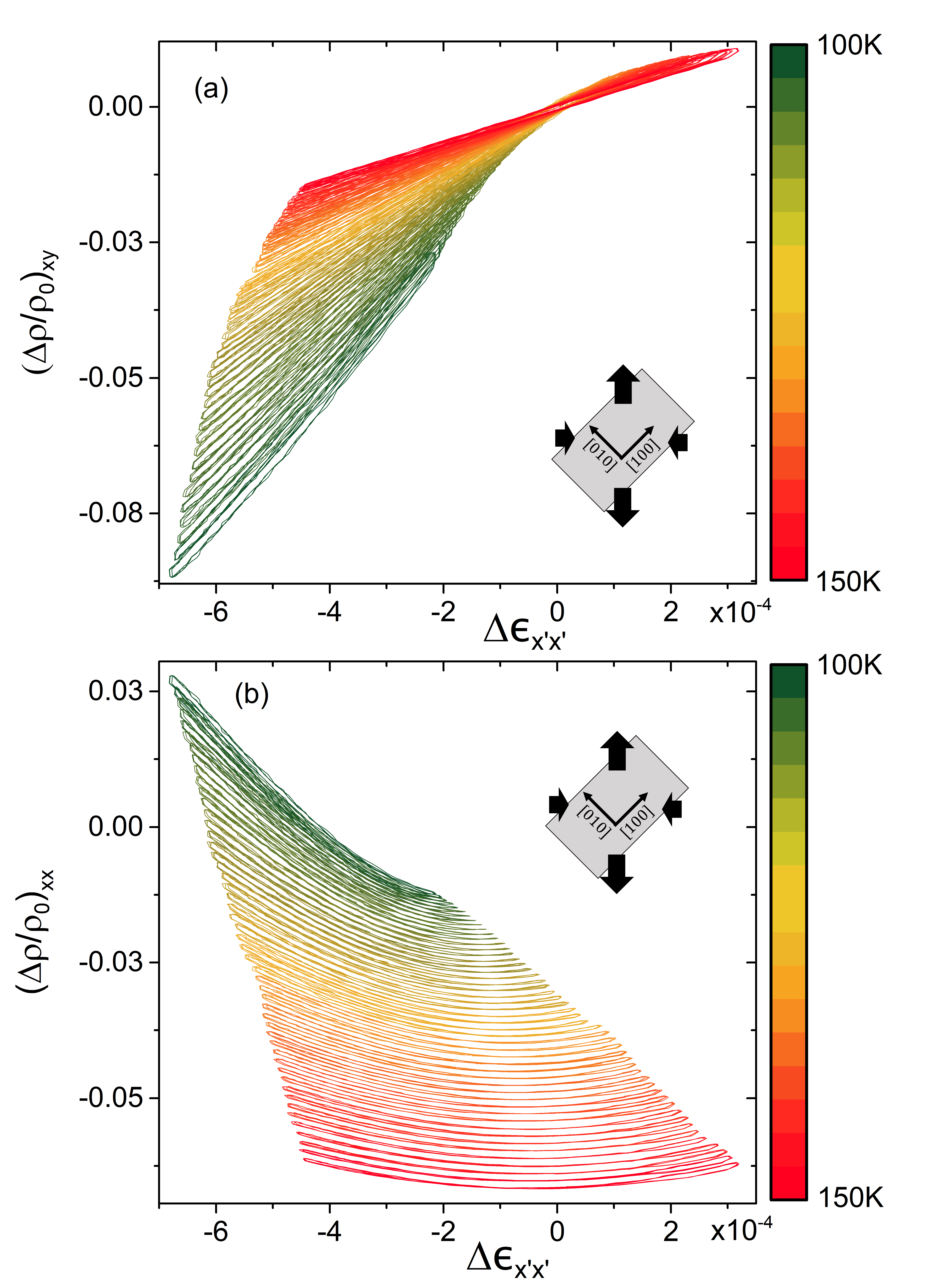}
\caption{\small \sl Temperature dependence of fixed temperature strain sweeps on Ba(Fe$_{0.975}$Co$_{0.025}$)$_{2}$A$_2$ under $A_{1g}$ and $B_{2g}$ symmetry strains measured using the $45^o$ measurement configuration. (a) the anisotropic resistivity response, $(\frac{\Delta\rho}{\rho_0})_{xy}$ (equivalent to $\frac{1}{2}[(\Delta\rho/\rho_0)_{x'x'}-(\Delta\rho/\rho_0)_{y'y'}]$ in the Modified Montgomery setup, shown in Fig. 2(a) of the main text), and (b) the isotropic elastoresistivity response $(\frac{\Delta\rho}{\rho_0})_{xx}$, (equivalent to $\frac{1}{2}[(\Delta\rho/\rho_0)_{x'x'}+(\Delta\rho/\rho_0)_{y'y'}]$ in the Modified Montgomery setup, shown in Fig. 2(b) of the main text). The small hysteresis is an experimental artifact and is discussed in the text. For clarity, the isotropic response fixed temperature strain sweeps are offset by $-1.5*10^{-3}$ from the 100K sweep. The anisotropic response is not offset. The trends shown in the $45^o$ measurement are consistent with the Modified Montgomery method measurements: the anisotropic resisitivity response becomes large closer to the structural transition, but is linear at all temperatures and the isotropic resistivity response has a large quadratic component that increases closer to the structural transition.}

\label{fig:Waterfall_Transverse}  
\end{center}  
\end{figure}

There are two important considerations when measuring nonlinear  elastoresistivity tensor components. First is the ability to accurately identify the neutral anisotropic strain point. The second is to simultaneously be able to measure $(\Delta\rho/\rho_0)_{A_{1g}}$ and $(\Delta\rho/\rho_0)_{B_{2g}/B_{1g}}$ for identical strain conditions. This has been demonstrated in the main text for the Modified Montgomery method \cite{Kuo2016, Shapiro2016a}. Another method that fulfills these requirements is the $45^o$ configuration, which is described in detail in Ref. \cite{Shapiro2016a}. This measurement setup allows the simultaneous measurement of $(\Delta\rho/\rho_0)_{xy}$ and $(\Delta\rho/\rho_0)_{xx}$ which corresponds to $(\Delta\rho/\rho_0)_{B_{2g}}$ and $(\Delta\rho/\rho_0)_{A_{1g}}$ repectively (or the measurement of $\frac{1}{2}[(\Delta\rho/\rho_0)_{x'x'}-(\Delta\rho/\rho_0)_{y'y'}]$ and $\frac{1}{2}[(\Delta\rho/\rho_0)_{x'x'}+(\Delta\rho/\rho_0)_{y'y'}]$, respectively, in the Modified Montgomery method). The neutral anisotropic strain point is extracted from where $(\Delta\rho/\rho_0)_{xy}$ crosses zero assuming that the ratio of the longitudinal contamination from small contact misalignment in the transverse voltage is constant as a function of temperature.

The qualitative behavior measured using both measurement configurations is in good broad agreement. The temperature dependence of the elastoresistance response of Ba(Fe$_{0.975}$Co$_{0.025}$)$_{2}$As$_2$ measured using the $45^o$ method for a sample experiencing  $A_{1g}$ and $B_{2g}$ symmetry strains can be seen in Fig. \ref{fig:Waterfall_Transverse}. Like the Modified Montgomery measurement (shown in the main text Fig. 2), the anisotropic resistivity response, $(\Delta\rho/\rho_0)_{B_{2g}}$, is always linear with a slope that increases as the sample is cooled towards the structural transition. The isotropic resistivity response, $(\Delta\rho/\rho_0)_{A_{1g}}$, shows a large increasing quadratic response as the sample is cooled towards the structural transition.

\begin{figure}[h]  
\begin{center}  
\includegraphics[width=4in]{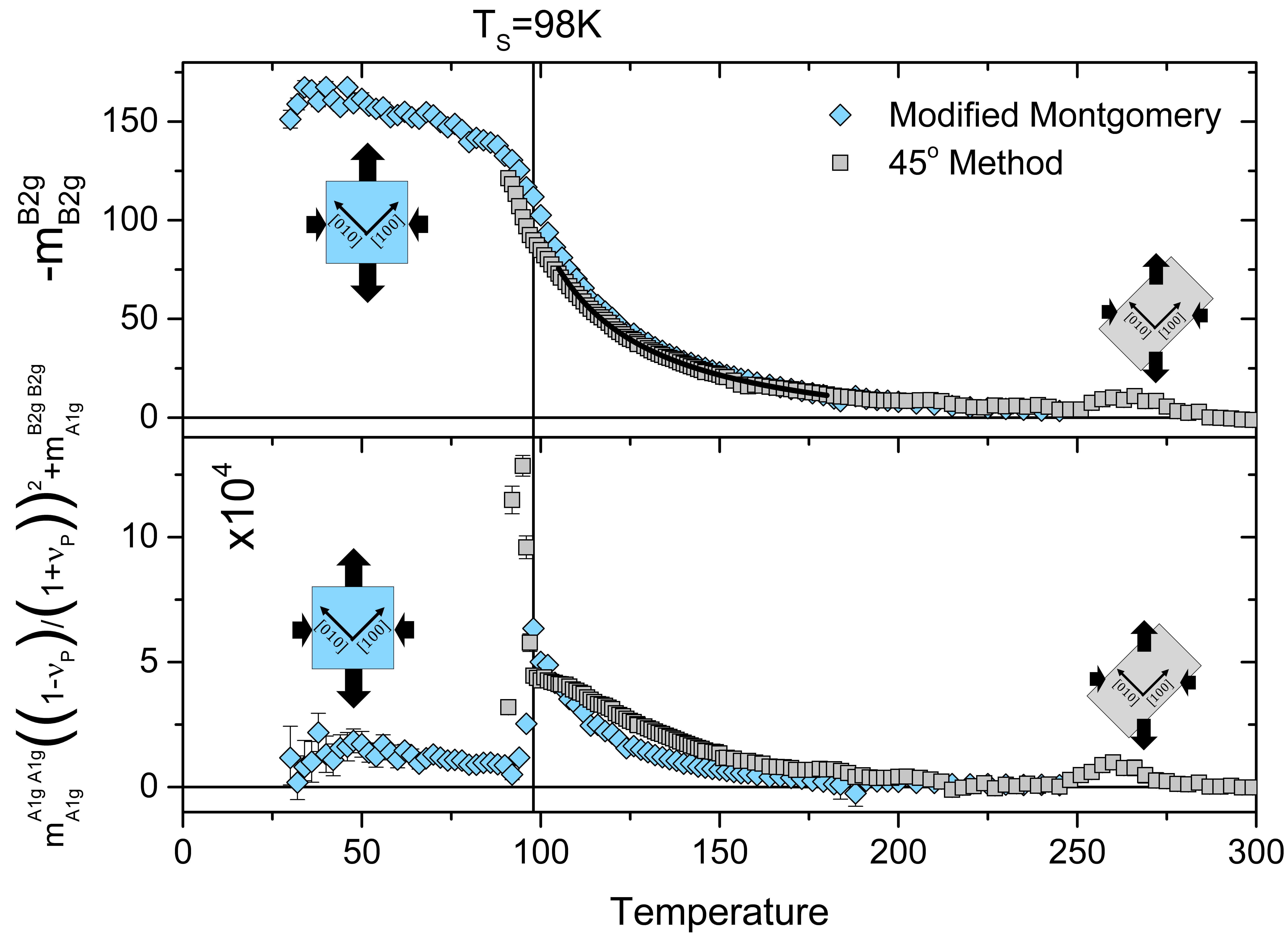}
\caption{\small \sl Comparison between the Modified Montgomery Method (light blue diamonds) and the $45^o$ setup (gray squares) for measuring elastoresistivity coefficients in Ba(Fe$_{0.975}$Co$_{0.025}$)$_{2}$A$_2$ under $A_{1g}$ and $B_{2g}$ symmetry strains. The Modified Montgomery sample shown is the one used in the main text. Both samples are in the size limit of maximal strain transmission and the data shown are not normalized. (a) The linear response to anisotropic strain, $m_{A_{1g}}^{B_{2g}, B_{2g}}$. Both methods can be fit by a Curie-Weiss temperature dependence (black line) with extracted Weiss temperatures of $\Theta=75.8 \pm 0.6 K$ and $\Theta=74.9 \pm 1.2 K$ for the Modified Montgomery sample and the $45^o$ method respectively. The Curie-Weiss fit for the Modified Montgomery sample can be seen in the main text Fig. 3(b). (b) The weighted quadratic coefficient in the isotropic resistivity response, dominated by $m_{A_{1g}}^{B_{2g}, B_{2g}}$. Both methods show qualitatively the same behavior; a large increase in the quadratic coefficient as the sample is cooled towards the structural transition. The functional form of the quadratic coefficient measured using the $45^o$ setup differs from the Modified Montgomery method, attributed to a small hysteresis in the strain sweeps and is discussed in more detail in the text.} 

\label{fig:Comp}  
\end{center}  
\end{figure}

This measurement suffers from the same experimental error as $B_{2g}$ Sample 2, shown in Fig. \ref{fig:ER_B2g}. The time constant during data acquisition was comparable to the time spacing between consecutive data points, introducing a small hysteresis in the fixed temperature strain sweeps. This can clearly be seen in Fig. \ref{fig:Waterfall_Transverse}.

The extracted elastoresistivity tensor components can be directly compared between the two techniques. Fig. \ref{fig:Comp}(a) shows the overlay of the fitted $m_{B_{2g}}^{B_{2g}}$ coefficients. The samples used for both measurements are in the size limit of maximal strain transmission ($\geq 80\%$), so the responses have not been normalized. Both measurements show a Curie-Weiss like divergence of the nematic susceptibility with a extracted Weiss temperature of $\Theta=75.8 \pm 0.6 K$ and $\Theta=74.9 \pm 1.2 K$ for the Modified Montgomery sample and the $45^o$ method respectively. The extracted quadratic coefficient, $((\frac{1-\nu_P}{1+\nu_P})^2m_{A_{1g}}^{A_{1g}, A_{1g}} + m_{A_{1g}}^{B_{2g}, B_{2g}})$, increases in both data sets upon approach to the structural transition. The data from the 45 degree measurement do not permit a more careful analysis of the temperature dependence or functional form. This is not intrinsic to the 45 degree measurements, and the quality of the data can likely be improved by setting the time constant of lock-in amplifier to appropriate values. Further measurements are underway to confirm this.

\begin{table}[h]
\begin{center}
\renewcommand{\arraystretch}{1.5} 
\begin{tabular}{m{23mm} m{16mm} m{20mm} m{16mm} m{12mm}}

\hline
\hline
\multicolumn{1}{c}{Method}  & \multicolumn{1}{c}{$m^{B_{2g}}_{B_{2g},0}$}  & \multicolumn{1}{c}{$\lambda/a_0$} & \multicolumn{1}{c}{$\Theta$ (K) } & \multicolumn{1}{c}{$R^2_{Adj}$}\\
\hline
 Modified Montgomery  & $17.3 \pm 0.7$ & $-2980 \pm 71$ & $75.8 \pm 0.6$ & $0.9995$ \\ 

$45^o$ setup & $14.5 \pm 0.9$ & $-2713 \pm 112$ & $74.9 \pm 1.2$ & $0.996$\\ 
 
\hline
\hline

\end{tabular}
\end{center}
\caption{\small \sl The fitted Curie-Weiss parameters and goodness of fit for fits of the linear anisotropic elastoresistivity response $m_{B_{2g}}^{B_{2g}}$ between 104K to 181K for Ba(Fe$_{0.975}$Co$_{0.025}$)$_{2}$A$_2$ under $A_{1g}+B_{2g}$ symmetry strains. Two fits are shown, one for a sample measured with the Modified Montgomery method and one for a sample measured with the $45^o$ setup. }
\label{table:fits} 
\end{table}

\end{document}